\documentclass[preprint]{aastex}

\usepackage{graphicx}
\usepackage{color}

\newcommand{\kmps}{km\,s$^{-1}$}
\newcommand{\cgsbunit}{erg\,s$^{-1}$\,cm$^{-2}$\,arcsec$^{-2}$}

\slugcomment{} \shorttitle{} \shortauthors{}

\begin{document}

\title{RW Aur A from the X-Wind Point of View: General Features}
\author{Chun-Fan Liu\altaffilmark{1,2} and Hsien Shang\altaffilmark{2,3}}
\affil{Institute of Astronomy and Astrophysics, Academia Sinica,\\
P. O. Box 23-141, Taipei 10641, Taiwan}

\altaffiltext{1}{Graduate Institute of Astrophysics, 
National Taiwan University, Taipei 10617, Taiwan}
\altaffiltext{2}{Theoretical Institute for Advanced Research in Astrophysics
(TIARA), Academia Sinica, P. O. Box 23-141, Taipei 10641, Taiwan}
\altaffiltext{3}{Send correspondence to: shang@asiaa.sinica.edu.tw.}

\begin{abstract}

In this paper, the RW Aur A microjet is studied from the point of view of X-wind models. The archived {\it HST}/STIS spectra of optical forbidden lines [O~{\sc i}], [S~{\sc ii}], and [N~{\sc ii}] from RW Aur A, taken in Cycle 8 with seven parallel slits along the jet axis, spaced at 0\farcs07 apart, were analyzed. Images, position-velocity diagrams, and line ratios among the species were constructed, and compared with synthetic observations generated by selected solutions of the X-wind. Prominent features arising in a steady-state X-wind could be identified within the convolved images, full-widths at half maxima and high-velocity peaks on both of the redshifted and blueshifted jets. The well-known asymmetric velocity profiles of the opposite jets are built into the selected models. We discuss model selections within the existing uncertainties of stellar parameters and inclination angle of the system. In this framework, the mass-loss rates that were inferred to be decreasing along the jet axis in the literature are the results of slowly decreasing excitation conditions and electron density profiles. Despite the apparent asymmetry in terminal velocities, line intensities and mass-loss rates, the average linear momenta from the opposite sides of the jet are actually balanced. These previously hard-to-explain features of the asymmetric RW Aur A jet system now find a different but self-consistent interpretation within the X-wind framework.

\end{abstract}

\section{Introduction}

The RW Aur system is one of the brightest and most studied members of the Taurus-Auriga star forming region \citep[$V\approx10.51$;][]{KH95}. It is a wide binary system with a separation of $\sim 1\farcs40$ \citep{WG01}. Both members possess circumstellar disks which are tidally interacting with each other \citep{CPPD06}. The primary, RW Aur A, is the only source that drives a bipolar jet. The jet was first inferred from its optical forbidden line profiles \citep{Ham94,HEG95} and later confirmed by narrow-band imaging \citep[HH 229;][]{ME98}. RW Aur A is known to be actively accreting with an accretion rate of $0.5 - 1.6\times10^{-6}$ $M_\odot$\,yr$^{-1}$ inferred from optical veiling \citep[e.g.,][]{HEG95,WH04}.

Overall, RW Aur A jet has a strikingly well-collimated appearance \citep[e.g.,][]{ME98,Dou00,Woi02}. The [S {\sc ii}] $\lambda\lambda6716+6731$ image of \citet{ME98} showed that the blueshifted jet lies exactly along the position angle of 130\degr\ at least up to 110\arcsec. The jet is marginally resolved within 56 AU by CFHT adaptive optics and {\it HST}/STIS. Widths of the jet slowly increase beyond this distance with an opening angle of 3\degr\ to 4\degr\ \citep{Dou00,Woi02}, and they appear to expand more rapidly closer to the source (within 0\farcs5) than farther away. Such trends of jet widths have been difficult to interpret in the past \citep{RB01}, as often the observed widths were interpreted as the extent of bow shock wings and the true jet should be much narrower. Farther than 0\farcs5 from the sources, the widths of the RW Aur A jet become wider as the intensity decreases along the jet axis, which was also observed by {\it HST} for HH 34 \citep{Ray96}. 
The fluxes of both blueshifted and redshifted jets decrease by an order of magnitude from 0\farcs5 to within 2\arcsec\ from the star. \citet{Woi02} considered the measured widths marginally consistent with an X-wind jet in \citet{SGSL} and the self-similar disk wind model in \citet{GCFB01} at high accretion efficiencies. 

Kinematics based on long-slit spectroscopy makes the RW Aur A jet a textbook example of a microjet. The position-velocity diagrams of its brightest forbidden lines, such as [S {\sc ii}], [O {\sc i}], and [N {\sc ii}], clearly display one single intensity peak at each position along the jet axis. The centroid velocity lies around 100 \kmps\ which is identified as the high-velocity gas tracing the collimated jet. The velocity widths are broad close to the source and narrow to its terminal velocities far along the length of the jet \citep[e.g.,][]{BHN96,HMS97,Pyo06,Mel09,HarHill09}. The low-velocity intensity peaks around $5$ to $20$ \kmps\ \citep{HMS97}, which are usually present in many classical T-Tauri stars, are virtually non-existent either in the optical {\it HST}/STIS data (\citealt{Woi02,Mel09}; and this work) or in the near-infrared [Fe {\sc ii}] $\lambda$1.644 \micron\ spectra from Subaru/IRCS \citep{Pyo06} and Keck/NIRSPEC \citep{HarHill09}. Although \citet{HEG95} and \citet{ABHC05} reported detections of some low-velocity peaks in [O {\sc i}] $\lambda6300$, these peaks were highly variable and have been discarded as simply transient on a timescale of months. The strong collimated emission at high velocity makes the RW Aur A jet one of the ideal candidates to confront with models of magnetocentrifugally-driven winds arising from a narrow region close to the inner-edge of the disks, for example, those that may resemble the X-winds \citep{SSG,XWindPPIV,XWindPPV}. In fact, it is our primary goal to compare this set of data with the best fits generated by models based on the X-wind solutions using approaches outlined in \citet{SGSL}.

Despite the predominant high-velocity emission, asymmetry across the opposite sides of the RW Aur A jet has long been documented in the literature \citep{HMSR94,HMS97,BHN96,Woi02,LCD03,Pyo06,Mel09,HarHill09}. The level of asymmetry in RW Aur A is among the largest in the 8 asymmetric T-Tauri jets reported in \citet{HMSR94}, who excluded interpretations based on different spatial orientations on opposite sides of the source, uncorrelated time-variabilities, different degrees of deceleration, or different degrees of collimation. Instead, they concluded that such strong asymmetry should arise in the acceleration process of the jet itself. \citet{Woi02} derived a velocity ratio of $1.6-1.76$ within 0\farcs5 from the source with the {\it HST}/STIS, and attributed the origin of asymmetry to the ``central engine''. \citet{LCD03} measured proper motions of the knots and found that they have the same ratios as their radial velocities, and the velocity asymmetry should be intrinsic to the jet. The ratios of proper motion are maintained well up to 4\farcm6 and 1\farcm6 from the source, along the blueshifted and redshifted jets, respectively \citep{MRF07}. We will treat the velocity asymmetry as an intrinsic property built in our models to explain the overall phenomena.

The asymmetry exists not only in flow velocities but also in line intensities and physical conditions derived from the line ratios. 
\citet{ME98} noted that the two opposite sides of the jet differ significantly in their spatial extent, with the [S {\sc ii}] jet extending to $\sim106\arcsec$ and $\sim50\arcsec$ on the blueshifted and redshifted sides, respectively. The red jet is brighter in the inner $\sim10\arcsec$ \citep{BHN96}, and remains so down to 0\farcs2 from the source as seen in both optical \citep{Mel09} and near-infrared \citep{HarHill09}. \citet{HMS97} noted that, at $\sim 1\arcsec$ angular resolution, the red jet has a larger electron density, a higher [S {\sc ii}] $\lambda6731$/[O {\sc i}] $\lambda6300$ ratio, and a lower H$\alpha$/[S {\sc ii}] $\lambda\lambda6716+6731$ ratio, than the blue jet. On the other hand, the faintness of the blue jet prevented \citet{BHN96} from attempting a reliable diagnostic analysis using ground-based spectra, which led \citet{Woi02} to assume similar [N {\sc ii}] $\lambda6583$/[O {\sc i}] $\lambda6300$ and [S {\sc ii}] $\lambda6731$/[O {\sc i}] $\lambda6300$ ratios for both sides of the jet for the {\it HST}/STIS spectra in their later analysis. \citet{Mel09} re-examined the same STIS dataset and obtained a smaller [N {\sc ii}] $\lambda\lambda6548+6583$/[O {\sc i}] $\lambda\lambda6300+6363$ but a larger [S {\sc ii}] $\lambda\lambda6716+6731$/[O {\sc i}] $\lambda\lambda6300+6363$ in the red jet, which imply a lower ionization fraction and temperature than the blue jet. We will re-examine the dataset, and analyze the line ratios and physical conditions based on the diagnostic approach developed in \citet{SGSL} for the X-wind jets.

As a result, the fundamental physical properties of the RW Aur A system may appear asymmetric. Mass-loss and momentum rates will be asymmetric when they are inferred from the measured asymmetric radial velocities, jet widths, and the electron densities obtained from line ratios \citep{BHN96,Woi02,Mel09}. \citet{Woi02} reported that if similar physical conditions had been adopted, the red jet would carry one half of the mass loss and one third of the momentum rates  carried in the blue jet. The imbalance of momentum may cause dynamical effects, such as recoil.
\citet{Woi02} calculated an upper bound 8\,km\,s$^{-1}$ for the recoil, under the assumption that the imbalance persisted during all the pre-main sequence lifetime of the star. They stated that this recoil value is probably a gross overestimate, because it is possible that earlier accretion of mass into the star might have been more symmetric, or that linear momentum balance might have been maintained over time. The dynamical time of these jets ($\sim10^5$ years or less) is relatively short, diminishing the reasons to expect an imbalance persistent during the total of the pre-main sequence phase (about $10^6$ years).
\citet{Mel09} revised the calculation by independently deriving electron densities and ionization fractions from each side, and derived similar mass-loss rates for the two sides. They attributed the asymmetry to inhomogeneous ambient material or to the magnetic configurations. 
Either work, \citet{Woi02} or \citet{Mel09}, leaves the momentum unbalanced in the presence of the observed velocity asymmetry. In this paper, we will undertake similar analyses based on the same {\it HST}/STIS dataset used in previous works, and demonstrate that the basic jet properties should be balanced in spite of the observed asymmetries in velocity and mass loss.

The X-wind framework provides an attractive possibility for models confronting with high-angular resolution observations. Through the star-disk interaction, the stellar magnetic fields can truncate the disk at the co-rotation radius in steady state, where the field is trapped and pinched toward a small area called the X-region at the inner edge of the disk. The material from the disk can accrete onto the star via a funnel flow along the field lines connecting the inner disk edge and the star. Meanwhile, the excess angular momentum from the star can be transported to the X-region and material can be magnetocentrifugally ejected along the open field lines as the X-wind. Streamlines of the wind flow fan out from the X-region, and gradually collimate along the axis of rotation. \citet{SSG} demonstrated the kinematic features of an X-wind through synthetic spectra with broad line widths close to the base of the jet which narrow to a profile of single velocity component upon approaching flow terminal speed. \citet{SGSL} self-consistently calculated the ionization and thermal conditions for semi-analytic X-wind solutions, and demonstrated how the excitation conditions can influence the synthetic images as ``optical illusions'', finding a good match of line ratios from available Herbig-Haro objects and microjets. Based on the works established in \citep{SSG,SGSL}, here we attempt a comparison of the X-wind models with those well-known features of the RW Aur A jet.

In this paper, we demonstrate a step by step comparison of the high-angular resolution data with theoretical models generated within the X-wind solutions through the analysis of key observed features. RW Aur A is an excellent candidate for such purposes due to its characteristic single high-velocity feature that simplifies the modeling. In Section 2 the analysis of {\it HST}/STIS archival data is addressed, and our parallel procedures for producing synthetic observable quantities from the models are presented. The observed properties of the jet extracted from the data: i.e., images, position-velocity diagrams, and the line ratios are discussed in \S\,3.1. An illustrative fit to the RW Aur A jet is given in \S\,3.2, and directly compared with the observational data. Finally the interpretation of these unusual features of RW Aur A are discussed in the theoretical framework based on the X-winds in Section 4.

\section{Methods}

\subsection{Archival \textit{HST}/STIS Data}

We obtained {\it HST}/STIS spectra of RW Aur A from the Multimission Archive at 
Space Telescope (MAST). The spectra were taken on 2000 December 10 and have 
been previously reported by \citet{Woi02,Woi05} and \citet{Mel09}. 
In this set of data, a $52\arcsec \times 0\farcs1$ slit was used with the 
{\tt G750M} grating. The slit was displaced at a step of 0\farcs07 parallel to 
the jet axis at seven different positions, creating a total field of view of 
about 0\farcs5 across the jet. Each exposure produced a two-dimensional spectral 
image of 52\arcsec\ long in the spatial axis and 6295 to 6867 \AA\ in the dispersion axis, 
covering the seven brightest emission lines: [O {\sc i}] $\lambda\lambda6300$, $6363$, 
[N {\sc ii}] $\lambda\lambda6548$, $6583$, [S {\sc ii}] $\lambda\lambda6716$, $6731$, 
and H$\alpha$. The slit sampled the emission at a nominal rate of 0\farcs05 pixel$^{-1}$ 
with an angular resolution of $\sim 0\farcs1$ (FWHM), and a spectral sampling rate 
of 0.554 \AA\ pixel$^{-1}$ with a 2.5-pixel line width, corresponding to $\sim25$ 
\kmps\ per pixel and a FWHM $\sim 65$ \kmps\ at $\sim 6500$ \AA.

The pipeline-processed spectra taken directly from MAST work reasonably well for our purposes. 
Hot pixels were present at specific positions on the CCD, but they fall primarily on 
line-free regions. Further reductions were performed to remove the baseline undulations 
caused by the wavelength rectification of undersampled stellar continuum and the contribution 
from the reflection nebula. Each of the seven spectral images was first divided into 
three sub-images containing lines of [O {\sc i}], H$\alpha+$[N {\sc ii}], 
and [S {\sc ii}], respectively. On each sub-image, we examined the rows containing 
the redshifted jet extending up to 4\arcsec\ and blueshifted jet up to 2\arcsec\ 
from the star. For each row, we fit the baseline by a Legendre polynomial, up to the 
10$^{\rm th}$ order, with IRAF/STSDAS\footnote{IRAF is distributed by the National Optical 
Astronomy Observatory, which is operated by the Association of Universities for 
Research in Astronomy (AURA) under cooperative agreement with the National Science 
Foundation; STSDAS is a product of the Space Telescope Science Institute, which is 
operated by AURA for NASA.} {\tt GFIT1D} and downhill simplex $\chi^2$ minimization.

Position-velocity (PV) diagrams were obtained for each line and each slit, relative to 
the systemic velocity of $+16.0$ \kmps\ \citep{Woi02}, and their transversely 
averaged PV diagrams. [O {\sc i}] $\lambda6300$ emission from velocity higher than 
$-235$ \kmps\ from the blueshifted jet was cut off due to wavelength coverage of the 
{\tt G750M} grating. Line profiles were also extracted from the PV diagrams every 0\farcs05. 
Each profile was fit with a single Gaussian and a linear baseline in IRAF/STSDAS 
{\tt NGAUSSFIT} with recursive $\chi^2$ minimization. On the intensity maps as 
constructed in \citet{Woi02}, a single Gaussian was fit to the transverse 
profiles at every 0\farcs05 with the Levenberg-Marquardt method for the amplitudes, 
intensity centroids, and full width at half maximum. For each pixel of the 2D STIS 
stack, ratios between pairs of lines were used to construct line-ratio diagrams for 
[S {\sc ii}] $\lambda\lambda6716/6731$, [N {\sc ii}] $\lambda6583$/[O {\sc i}] $\lambda6300$, 
and [S {\sc ii}] $\lambda6731$/[O {\sc i}] $\lambda6300$. The data points were further arranged according to their spatial distribution.

In the following analysis performed on the model data cubes, operations and conditions applied in the observational analysis were adapted so that the model dataset can be properly compared with the observed dataset.

\subsection{Synthetic Maps}

Synthetic images and long-slit spectra are calculated for various X-wind configurations with methods and examples illustrated in Shang et al.\ (1998, henceforth SSG) and Shang et al.\ (2002, henceforth SGSL).
We follow the approach adopted in SGSL to obtain a self-consistent thermal structure for each model, in which various ionization and recombination and heating and cooling processes associated with the wind are considered. These include hydrogen photoionization and H$^-$ photodetachment by photons from the star and accretion hot spots, X-ray heating and ionization, collisional ionization, ambipolar diffusion heating, and external heating arising from mechanical origins in the flows. In a magnetized star-disk system, the closed stellar magnetic loops confine plasma at soft X-ray temperature, and sites of magnetic field line reversals between the star and the wind are prone to reconnection events that can generate harder X-rays (see, e.g., Figure 1 of Shu et al.\ 1997). SGSL concluded that X-rays are effective in ionizing the base of the X-wind (see also Shang et al.\ 2010, henceforth SGLL), and mechanical heating can easily heat the flow to 8000-10000 K with a small amount of mechanical energy converted to heat. In this work, we adopt conventions defined in SSG and SGSL, and find solutions normalized by stellar parameters for RW Aur A system.

We extend the reaction network to include the ionization balance of oxygen, sulfur, and nitrogen.  Charge exchange with the H atom controls the equilibria between O$^0$ and O$^+$, and N$^0$ and N$^+$ as their ionization potentials are similar (13.60, 13.61, and 14.53 eV for H, O, and N, respectively). Sulfur is assumed to be ionized as a result of its lower ionization potential (10.36 eV). Atomic data from \citet{Men83} were used in these calculations, and updated collisional strengths of [S {\sc ii}] \citep{CP93} were adopted.

To obtain optimal fitting with the observed spectra, the parameters for mass-loss rate $\dot{M}_{\rm w}$ defined in \citet{XWindI}, X-ray ionization $L_{\rm X}$ (Equation 3.10 in SGSL), and coefficient of heating $\alpha_h$ (Equation 5.2 in SGSL) were adjusted to obtain proper thermal profiles for emission. The fiducial cases shown in SGSL and Shang et al.\ (2004, henceforth SLGS) were taken as initial model standards. 
The initial values of $\dot{M}_{\rm w}$ and $L_{\rm X}$ were first allowed to vary with the ratio $L_{\rm X}/\dot{M}_{\rm w} = 3\times10^{13}$ erg\,g$^{-1}$ fixed, a measure of X-ray attenuation previously in SGSL and SGLL.
The values of $\alpha_h$ are adjusted to produce thermal profiles that would generate emissions fluxes matching the observed ones. The properties of the synthetic images are derived and spectra for further analysis and comparisons with the STIS data. We note that the actual stellar system always looks knottier in emission than the synthetic profiles produced by the steady-state models, and that the level of X-rays adopted in our models in this paper is on the higher end of flares attained by low-mass pre-main sequence stars \citep[e.g.,][]{Get08,MW11}. As explained in SGLL, the higher level of X-ray luminosity adopted in our modeling scheme may reflect the fact that these X-rays may have experienced higher extinction near the base of the wind to reach the level of ionization obtainable for an otherwise lighter wind. A lower X-ray luminosity can be adopted instead if the X-rays are generated at a more elevated position relative to where the ionization actually takes place. This can effectively lower the absorption column in the context of X-wind models (unpublished work). By adopting and keeping the fixed ratio $L_{\rm X}/\dot{M}_{\rm w} = 3\times10^{13}$ erg\,g$^{-1}$, the analysis shown in this work is consistent and can be easily compared with past works using this approach (SGSL, SLGS, and SGLL). We define and focus on the overall trends and profiles rather than local variations within individual knots in the process of model-data comparison and fitting process.

The synthetic images have to be generated at very high-resolution using high sampling 
rates for convolution. The maps were convolved with the telescope point spread function 
while retaining spatial sampling in the original map. STIS instrumental 
profiles were applied to the long-slit spectra. Each of the profiles was convolved with a two-dimensional 
Gaussian of $0\farcs1 \times 65$ \kmps\ FWHM, and resampled at 0\farcs05 and 25 \kmps, 
based on profiles of STIS $52\arcsec \times 0\farcs1$ slit and {\tt G750M} grating. 
The resulting PV diagrams were fitted with a one-dimensional Gaussian 
at each spatial sampled point using the downhill simplex method. The fitted values of 
line intensities, velocity centroids, and velocity widths were compared with those derived 
from the observed spectra along the jet axis.

The synthetic line ratios of [S {\sc ii}] $\lambda\lambda6716/6731$, 
[S {\sc ii}] $\lambda6731$/[O {\sc i}] $\lambda6300$, and 
[N {\sc ii}] $\lambda6583$/[O {\sc i}] $\lambda6300$  were calculated diagnosing the jet 
from different positions in the integrated line intensity maps.

\subsection{Properties of the RW Aur A System}

Properties of the RW Aur A system affect the actual model fits to observations as the stellar parameters determine the actual dimensional values of the models. The systematic velocity also affects the values of the observed jet speeds relative to the driving source and the velocity ratios. Furthermore, the inclination angle may introduce uncertainties to the terminal velocities. Stellar parameters in the literature determined from various evolutionary tracks, veiling $r_\lambda$ and visual extinction $A_V$, stellar mass $M_\ast$, stellar radii $R_\ast$, spectral types and luminosity  are summarized in Table~\ref{starpars} \citep{HEG95,SBB00,WLK01,WG01}. It can be seen that the ranges of inferred stellar masses, radii and luminosities have not varied more than a factor of two. Within the uncertainties, we adopt a reference value of $M_\ast=1 M_\odot$ and $R_\ast=2 R_\odot$ for illustrative purposes throughout this work.

RW Aur A may have a varying heliocentric radial velocity. For the same STIS dataset, \citet{Woi02} adopted a value of $+16.0$ \kmps\ \citep{Pet01} and obtained an average velocity ratio for the jet components of $\sim 1.6$. On the other hand, \citet{Mel09} adopted $+23.5$ \kmps\ \citep{Woi05} and obtained a higher value of the ratio $\sim 1.8$. 
The value obtained by \citet{Woi05} was based on one single fitting to the Li~{\sc i}~$\lambda6708$ absorption in the current {\it HST}/STIS dataset, which we reanalyze with a value of $+22.1\pm0.61$ \kmps. For the illustrative purpose of this work, we adopt the averaged value of $+16.0$ \kmps\ \citep{Pet01,WH04} and set the nominal velocity ratio to be 1.6 throughout this work. The radial velocities reported in the literature are summarized and their inferred jet velocity ratios in Table \ref{RV}.

Various groups have attempted to derive the inclination angle of the RW Aur A system. \citet{Woi02} compared proper motions of the RW Aur redshifted jet between CFHT/PUEO (1998) and {\it HST}/STIS (2000) observations and derived a value of 53\degr. \citet{LCD03} improved the value to be 46\degr\ for both jets with corrected PUEO pixel size. \citet{ABHC05} modeled the emission and absorption profiles of Balmer lines and Na~{\sc i}\ D lines with a disk wind model of \citet{BP82_DiskWind} and inferred a range between 55\degr\ and 65\degr. \citet{CPPD06} modeled $^{12}$CO ($J=2-1$) spectra of the disk around RW Aur A with Keplerian disk profiles following \citet{BS93}, and found that 
the line profiles can be matched with 45\degr\ or 60\degr. The range of inclination angles agrees reasonably well within 45\degr\ and 60\degr, which we will take as the uncertainty introduced when one de-projects velocities. 

Within the range of stellar parameters, we explore combinations of wind mass-loss rate $\dot{M}_{\rm w}$, disk truncation radius $R_{\rm x}$ (the X-point, see \citealt{XWindI}), and solutions of the X-wind models (\citealt{XWindI,XWindII,XWindV}; \citealt{Jet_IAUS}; \citealt{ShangPhD}; \citealt{SSG,SGSL,SLGS,SGLL} [SSG, SGSL, SLGS, SGLL]; \citealt{CSLS08}). Details of the magnetic configurations and their observational properties will be presented in a separate publication. In this work, we focus on specific features that one can derive from the observed properties of the RW Aur A jet, which are applied as selection criteria for the models. The X-winds that satisfy the identified constraints are further analyzed and compared with the STIS dataset.

The velocity asymmetry, if intrinsic to the system, reveals an interesting connection of winds launched from the opposite sides of the disk in the X-wind context. In steady state, the inner disk truncation radius coincides with the X-point, and is co-rotating with the star at $\Omega_\ast=\Omega_{\rm x}\equiv {(GM_\ast/R^3_{\rm x})}^{1/2}$. This introduces the unit of velocity, $v_{\rm x}=R_{\rm x}\Omega_{\rm x}$, defined at the base of the jet \citep{XWindI,XWindII,XWindV}, where the velocity of the wind 
can be non-dimensionalized as $\bar{v}_{\rm w}=\bar{\Upsilon}_{B,R}\,v_{\rm x}$. The velocity ratio, $\bar{v}_{{\rm w},B}/\bar{v}_{{\rm w},R}\approx 1.6$, suggests the launch of unequal winds above and below the disk from the X-point. The $R_{\rm x}$ is set as one single uniform value for both sides of the disk around the circumference. For the scope of this paper, we do not consider detailed launch mechanisms that could introduce the asymmetric mass loading or dynamical effects which may warp 
the disk or cause a non-circular orbit at the innermost edge. Such an investigation shall be reserved for future studies.

We select X-wind solutions whose properties satisfy these conditions from \citet{ShangPhD}. The ratio of the averaged velocity between the blue and the red jet components yields
 $${\bar{v}_{{\rm w},B}\over\bar{v}_{{\rm w},R}} 
   = {{\bar{\Upsilon}_B\,v_{\rm x}}\over{\bar{\Upsilon}_R\,v_{\rm x}}}
   = {{\bar{\Upsilon}_B}\over{\bar{\Upsilon}_R}}.$$ 
Without specifying the actual value of $R_{\rm x}$, the selection of solutions will not be unique. 
In Tables~\ref{SetI} and \ref{SetII}, we give two sets of examples. The range of $R_{\rm x}$ can vary from a value smaller than 0.05 AU, to as large as 0.2 AU. Note that we have set no {\it a priori} condition that restricts the choices of $R_{\rm x}$ by either stellar or model parameters. The reality probably exists somewhere between the two extreme values. Although further constraints may be available from other clues of the system, we leave it as a subject in our followup investigations.

In Section 3, we analyze and compare with the archived {\it HST}/STIS data based on one choice of $R_{\rm x}$, one inclination angle, and one pair of solutions. The blue and the red sides are treated independently, although the two sides of the jet are connected through one common point $R_{\rm x}$ in space and the fixed ratio of averaged velocities. The equivalent individual mass-loss rates $\dot{M}_{\rm w}$ defined in the sense of the original X-wind parameters \citep[e.g.,][]{XWindI,XWindII} are adjusted to match the average intensities of knots in the respective side. The actual mass-loss rates that enter into either hemisphere are half of the inferred model values. The individual mass-loss rates for each hemisphere are reported in \S\,3 and \S\,4.

\section{Results}

In this section, an example to demonstrate the general principles in applying the X-wind models to observations is illustrated. We first show the observed features of the RW Aur A jet extracted from the archival {\it HST} data.

\subsection{Observed Features}

\begin{figure}
\epsscale{0.7}
\plotone{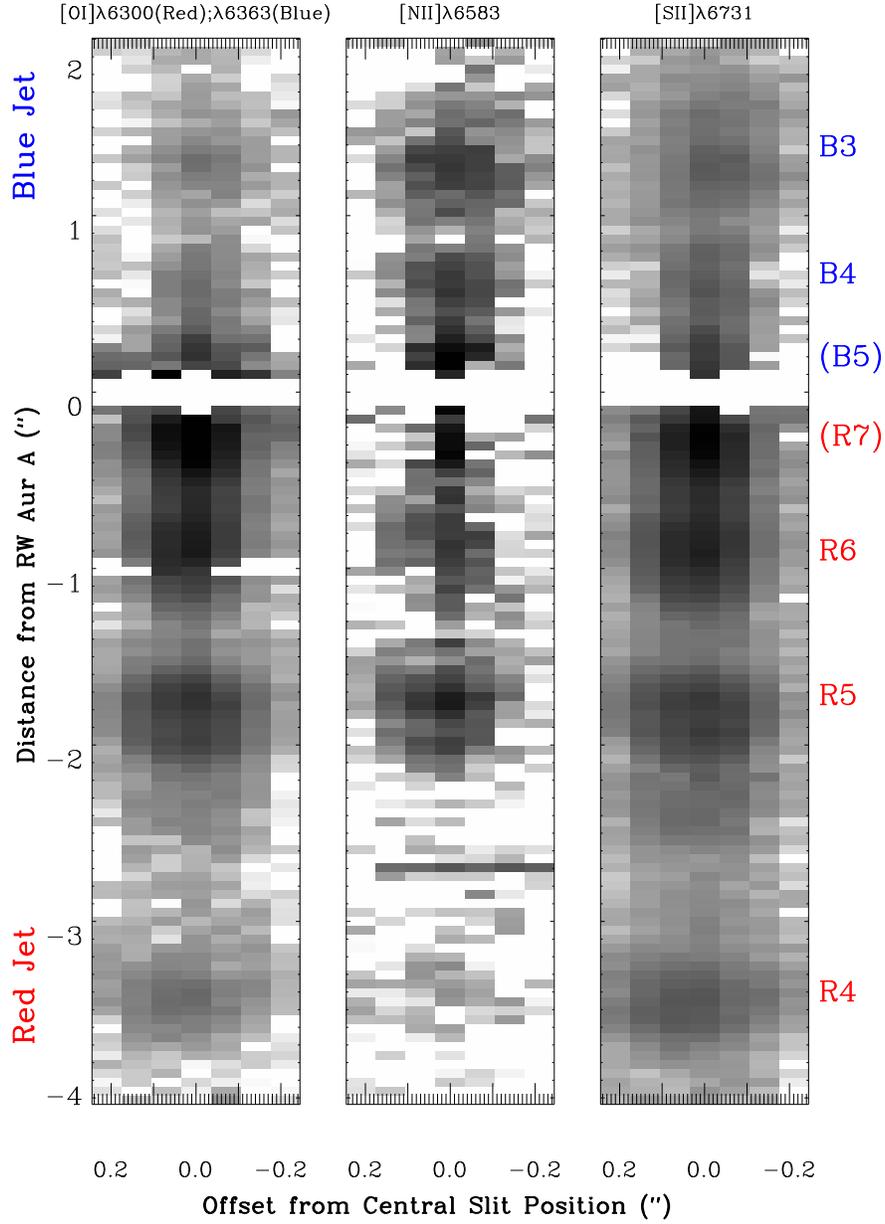}
\caption{Narrow band images of the RW Aur A jet extracted from the {\it HST}/STIS archive. The jet axis is plotted as the abscissa, with the distance increasing along the position angle of 130\degr. The ordinate is the distance across the jet and is marked by positions of the seven slits separated by 0\farcs07. The images are logarithmically scaled between $5\times10^{-16}$ and $5\times10^{-13}$ \cgsbunit. On each panel, the knots are identified and labeled as in \citet{LCD03}, and the ``stationary knot'' on each side of the jet is labelled in parentheses by this work.}
\epsscale{1.0}
\label{ObsImg}
\end{figure}

The two sides of the RW Aur A jet are asymmetric in appearance. The narrow-band images of 
three forbidden lines [S {\sc ii}] $\lambda6731$, [N {\sc ii}] $\lambda6583$, and [O {\sc i}] 
($\lambda6300$ for the red jet and $\lambda6363$ for the blue jet) are shown with gray scale in 
Figure \ref{ObsImg}. Both the blue and red jets 
are knotty, and the knots become fainter as the jet propagates
outwards. The innermost knots appear to be tightly connected while the distance between the 
outer knots increases. The red jet extends beyond 4\arcsec\ and at least four knots can 
be identified. The blue jet is overall dimmer by factors of 3 to 4, and extends up 
to 2\arcsec\ with three knots identified. The extent of asymmetry slightly differs among 
line species. [N {\sc ii}] is overall dimmer and shows less asymmetry than [O {\sc i}] 
and [S {\sc ii}], and lacks emission beyond R4. The red jet appears somewhat more compact in 
the transverse direction than the blue jet.

\begin{figure}
\epsscale{0.7}
\plotone{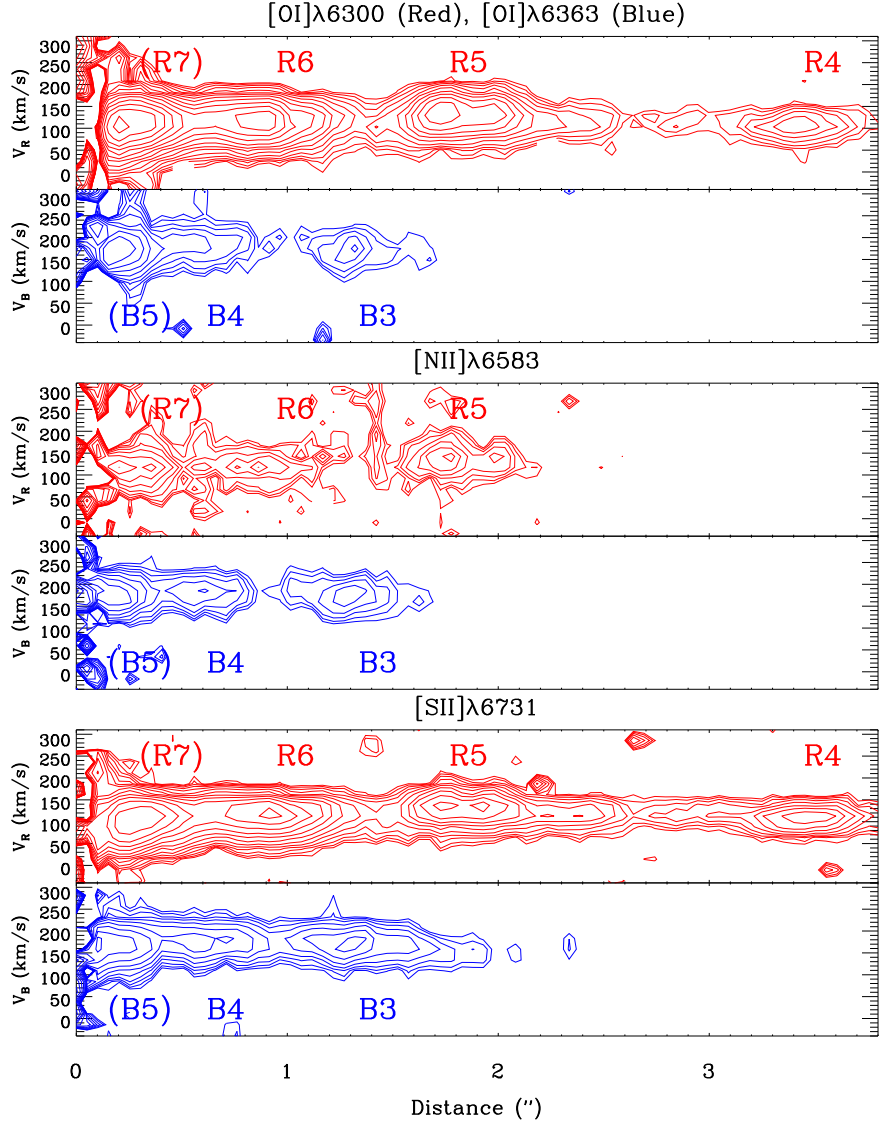}
\caption{Position-Velocity diagrams of the central 0\farcs1 in width of the RW Aur A jet derived from the {\it HST}/STIS spectra. $V_R$ and $V_B$ stand for the absolute values of the projected velocities of the red and the blue jet, respectively. Contours plotted start from $3\sigma$ and increase by $\sqrt{2}\sigma$. The $1\sigma$ of the lines are derived from the line-free regions next to the specific lines: $4.91\times10^{-16}$, $6.15\times10^{-16}$, $5.15\times10^{-16}$, and $4.96\times10^{-16}$ \cgsbunit\ for [O {\sc i}] $\lambda6300$, [O {\sc i}] $\lambda6363$, [N {\sc ii}] $\lambda6583$, and [S {\sc ii}] $\lambda6731$, respectively.}
\epsscale{1.0}
\label{ObsPV}
\end{figure}

The asymmetry in intensities and velocity centroids are indeed pronounced features on the PV diagrams. The velocity ratios between the blue and red jets are $\sim$ 1.56 to 1.60, which can be traced to the innermost detectable region of the system at 0\farcs2. 
Figure \ref{ObsPV} shows the PV diagrams of [O {\sc i}] ($\lambda6300$ for the red jet and $\lambda6363$ for the blue jet), [N {\sc ii}] $\lambda6583$, and [S {\sc ii}] $\lambda6731$ extracted from the central slit. The velocity ranges from $-25$ to $+300$ \kmps\ in the red, and from $-350$ to $+25$ \kmps\ in the blue. The knots were identified by peak positions in the PV diagrams and are labeled as in Figure \ref{ObsImg}. The line shapes of the knots can be fit with a single Gaussian profile throughout the jet, showing clearly that the jet is dominated by a single velocity component. The velocities at the centroids range between $-160$ and $-180$ \kmps\ and $+100$ and $+120$ \kmps, on the blueshifted and redshifted sides, respectively. Velocity centroids within the observed field of view vary slightly among the species. The velocity along the jet varies, as the knot R5 in the red jet is $\sim 10$ \kmps\ faster than the overall average, and the knot B3 in the blue jet shows a decreasing trend on the order of $\sim 20$ \kmps.

\begin{figure}
\plottwo{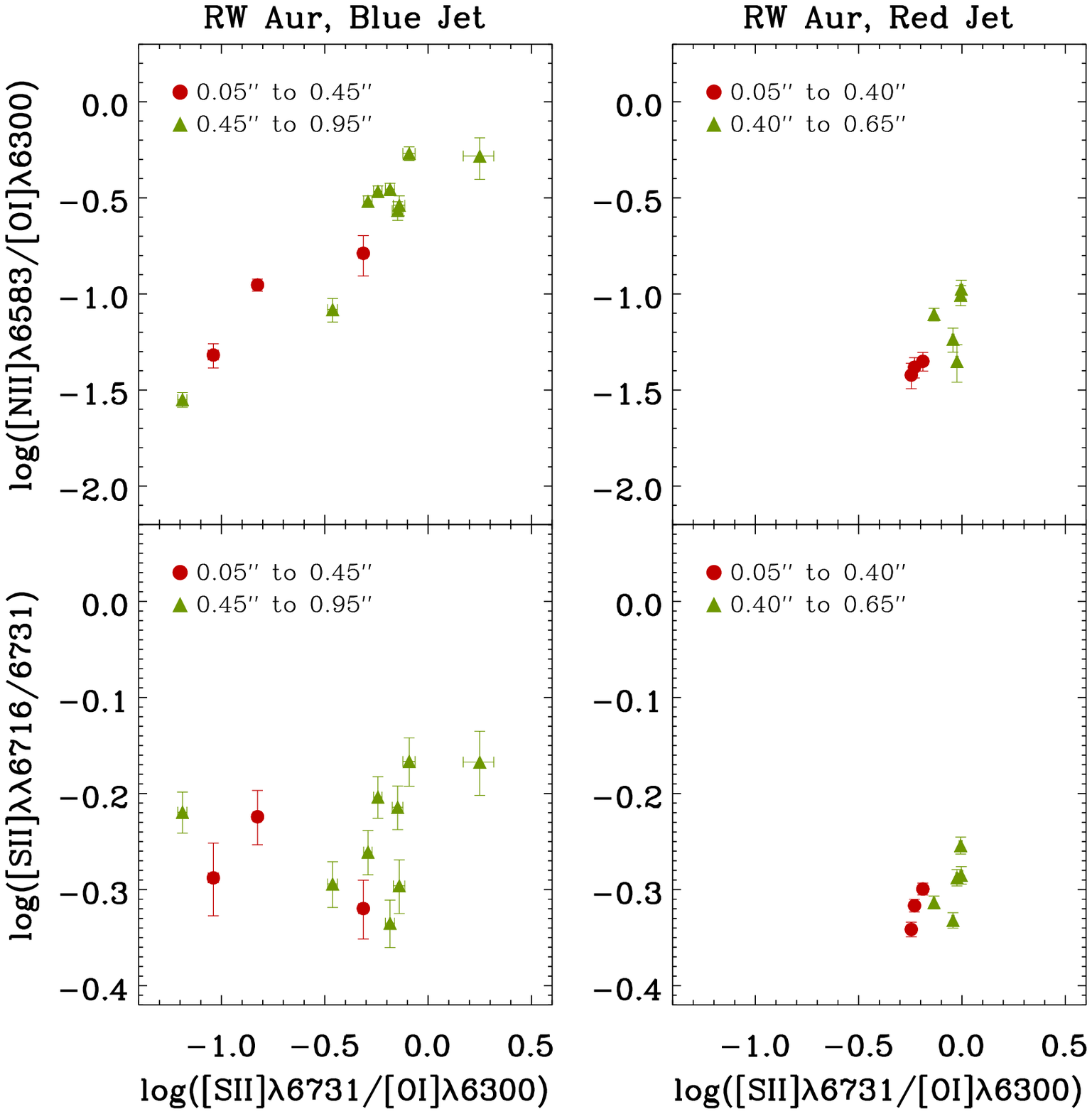}{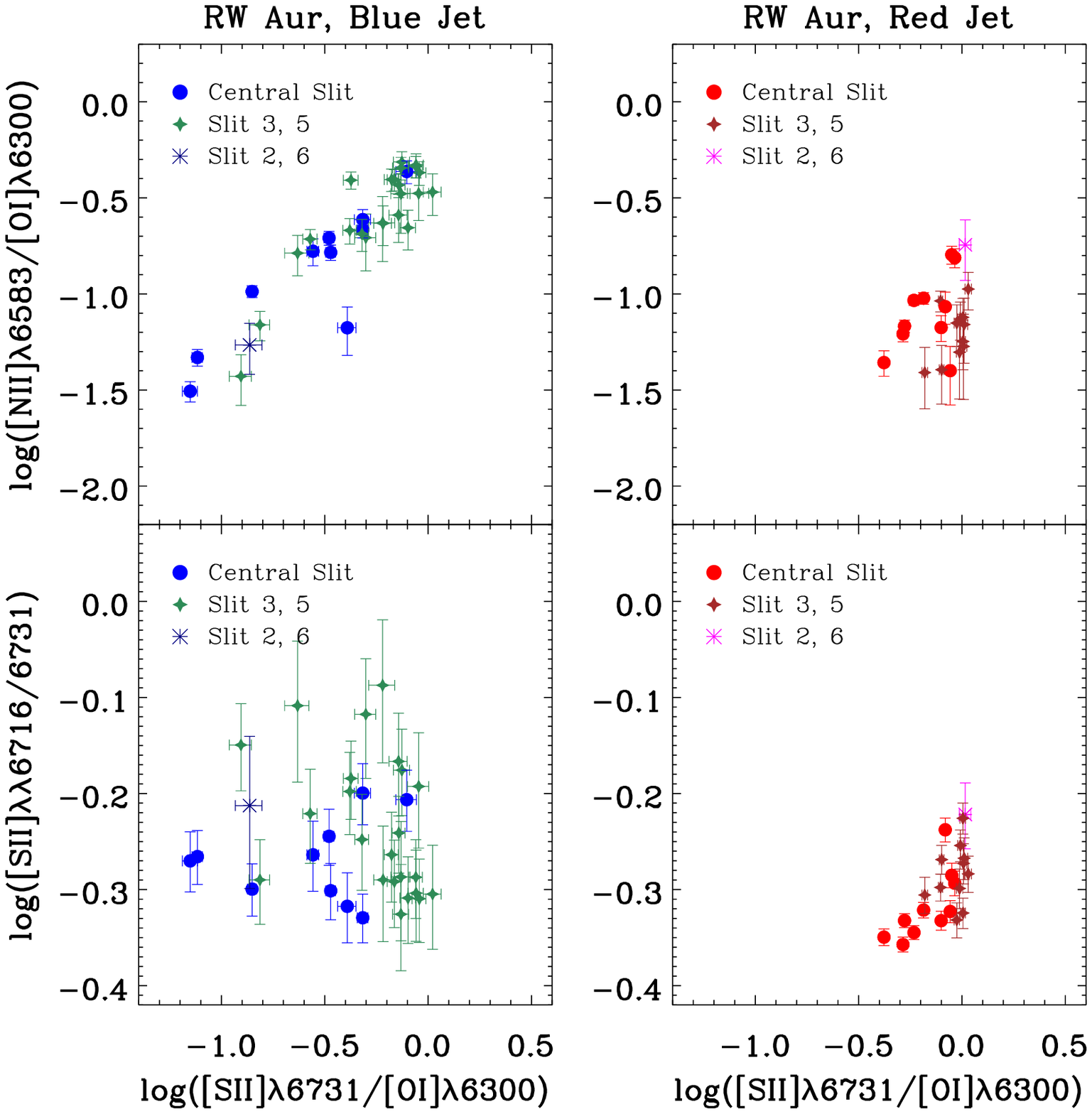}
\caption{Optical line ratio diagrams for the RW Aur A jet derived from the {\it HST}/STIS spectra. The blue jet is defined here by the integrated intensities over velocity interval [$-250,-20$] \kmps\ and the red jet is defined by those within [$+20,+200$] \kmps. On each panel, points labelled by the same color are sampled every 0\farcs05 unless the fractional error of either line ratio is above 50\%. {\it Left:} Line ratios integrated over the central five slits at different distances along the jet. {\it Right:} Line ratios obtained at different positions from the central five slits without integration. The innermost 0\farcs95 of the blue jet and 0\farcs65 of the red jet are shown for clarity.}
\label{ObsLR}
\end{figure}

Loci of line ratios from the blue and red jets do not overlap on the cross-ratio plots. The line ratios between [N {\sc ii}] $\lambda6583$/[O {\sc i}] $\lambda6300$ and
[S {\sc ii}] $\lambda6731$/[O {\sc i}] $\lambda6300$, and between [S {\sc ii}] $\lambda\lambda6716/6731$ and [S {\sc ii}] $\lambda6731$/[O {\sc i}] $\lambda6300$ are shown in Figure \ref{ObsLR}.  We show the line ratio plots for points within 0\farcs95 (B4$+$B5) from the source for the blue jet, and within 0\farcs65 (R7) for the red jet, at a distance ratio approximately proportional to their respective velocities. Line ratios integrated over slits labelled by their positions from the source are shown in the left panels, and those from the individual slits are shown in the right panels. 
Only those points with relative uncertainties lower than 50\% are shown. We focus on the analysis for data points from the three central slits since the outer slits do not contribute statistically significant information as seen in the right panel of Figure \ref{ObsLR}. The line ratios increase roughly with distance along the jet axis. The line ratios derived from different slits do not segregate into distinct groups, but rather scatter around a general trend. The emissions appear to come from continuous media of similar physical conditions within each side of the jet, but each component has its own distinct trends. In the red jet component, the [S {\sc ii}] $\lambda\lambda6716/6731$ ratio clusters around similar values $0.45-0.63$ whereas the [S {\sc ii}] $\lambda6731$/[O {\sc i}] $\lambda6300$ ratio ranges from 0.3 to 1.0. It also shows an increasing trend with increasing values of [S {\sc ii}]$\lambda6731$/[O {\sc i}]$\lambda6300$ from the outer slits. In the blue jet component, both the ratios of [S {\sc ii}] doublets and [S {\sc ii}] $\lambda6731$ to [O {\sc i}] $\lambda6300$ have large scatters due to the weakness of the lines, and there is no clear correlation of the values. Similar trends exist for the [N {\sc ii}] $\lambda6583$/[O {\sc i}] $\lambda6300$ ratio: those in the blue span a wider range while those in the red are located within a confined region, however, the two line ratios are clearly positively correlated, from the central to outer slits. Perhaps due to the faster drop-off of [O {\sc i}] signal particularly in the blue jet, points from farther distance tend to drift and pile upward in the upper-right corner of the [N {\sc ii}] $\lambda6583$/[O {\sc i}] $\lambda6300$ to [S {\sc ii}] $\lambda6731$/[O {\sc i}] $\lambda6300$ panel with even larger scatter. Those points were not displayed in Figure \ref{ObsLR} for clarity of presentation. 

\subsection{A Sample Fit for the RW Aur A System}

In this subsection, a sample fit for the RW Aur A jet is shown. For simplicity, we demonstrate one 
case that could capture overall the major features of the jet in this paper. Other 
solutions are possible, within the observational constraints set by the {\it HST}/STIS 
data, such as those listed in Table \ref{SetII}. 

The sample is described by properties outlined in Section 2.3 for the RW Aur A: stellar parameters ($M_\ast,R_\ast$) $=$ ($1M_\odot,2R_\odot$), a systemic velocity of $+16.0$ \kmps\ (Table \ref{RV}), an inclination angle of 55\degr\ (Table \ref{SetI}), and a disk truncation radius $R_{\rm x}=0.12$ AU (see Table \ref{SetI} for the range of possible choices). With these parameters, one solution gives an average projected velocity of $-170$ \kmps\ for the blue side and the other gives $+115$ \kmps\ for the red side as inferred from the archival data. The velocity coverage is in the range of 200 to 310 \kmps\ for the blue, and 120 and 200 \kmps\ for the red. They possess properties of those pointed out by SSG such as the cylindrically stratified density structures, broad line profiles near the bases, and slowly-collimating streamlines toward the axis over a large distance. The density contours and streamlines are similar to Figure 1, and the PV diagrams are similar to Figure 3 of SSG.

\begin{figure}
\plotone{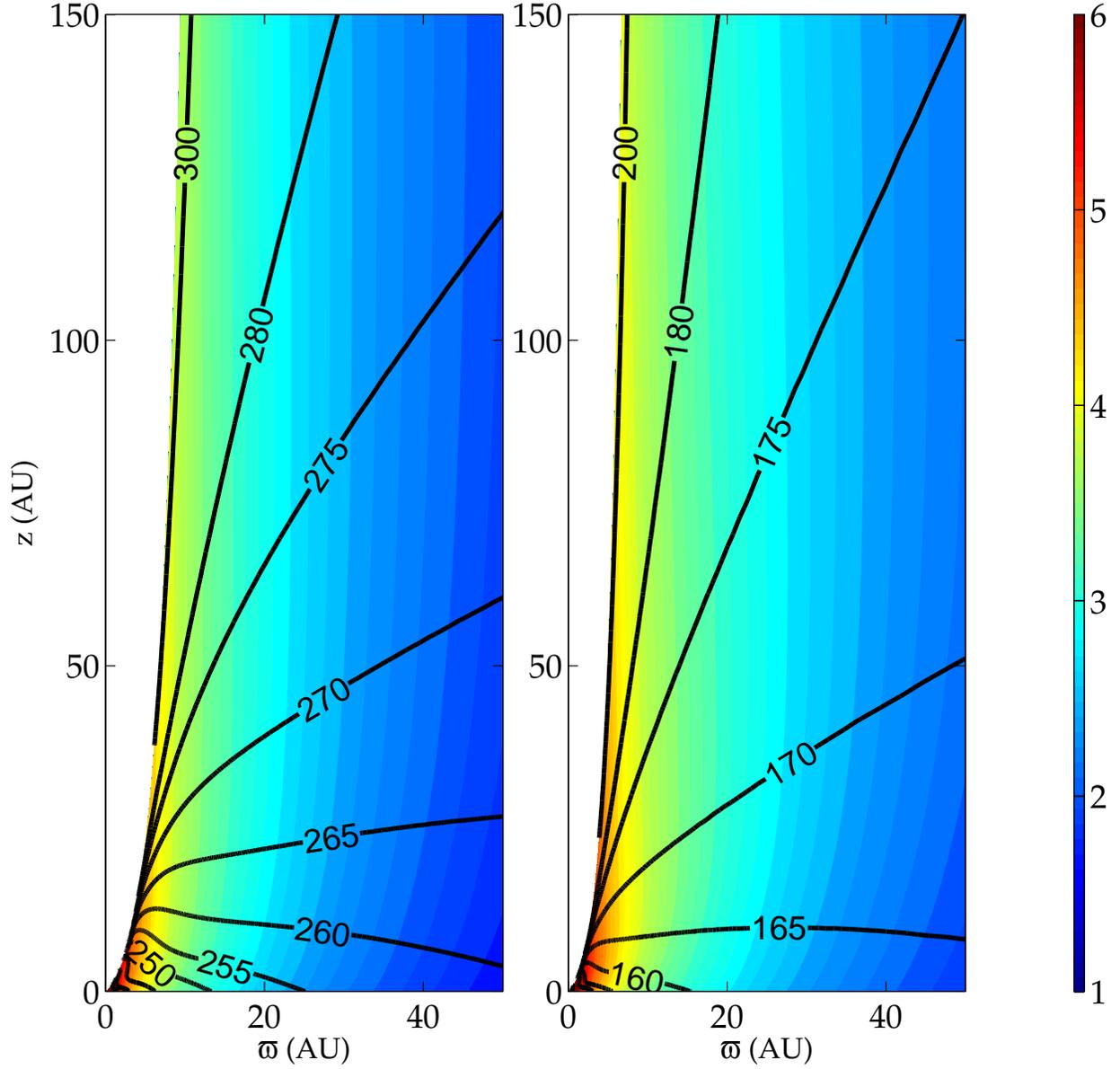}
\caption{Profiles of the velocity (black contours, in unit of \kmps) and electron density (color map, in logarithmic unit of cm$^{-3}$) of the X-wind solutions for the sample parameters listed in Table \ref{RunPars}. Models adopted for the blue and red jets are shown on the left and right, respectively.}
\label{NeVpPlot}
\end{figure}

\begin{figure}
\plotone{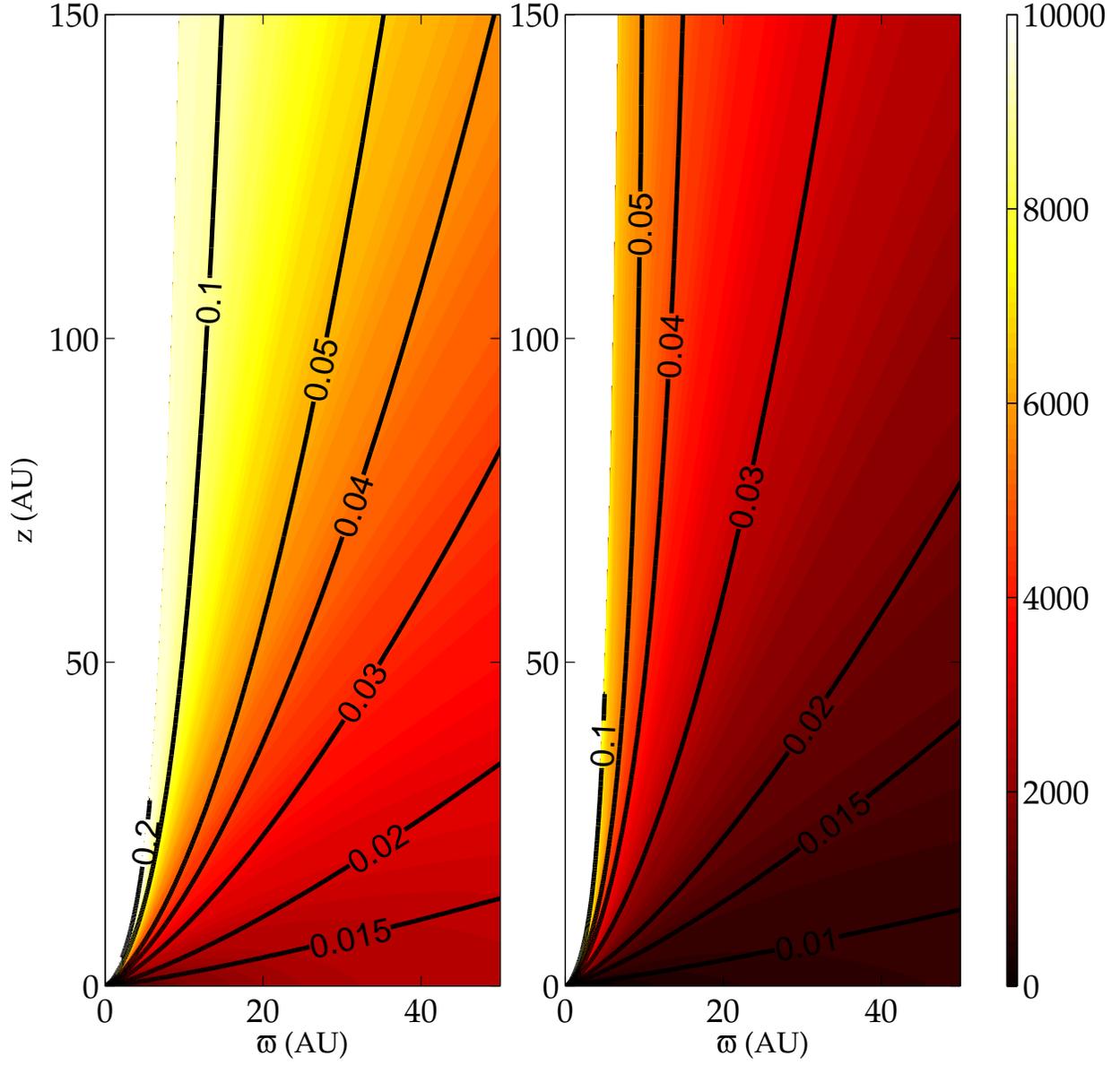}
\caption{Contours of electron fraction (black lines) are shown on top of the temperature profiles (color map, in unit of K), for the models listed in Table \ref{RunPars}. The left panel is for the blue jet and the right panel is for the red jet.}
\label{XeTPlot}
\end{figure}

\begin{figure}
\plottwo{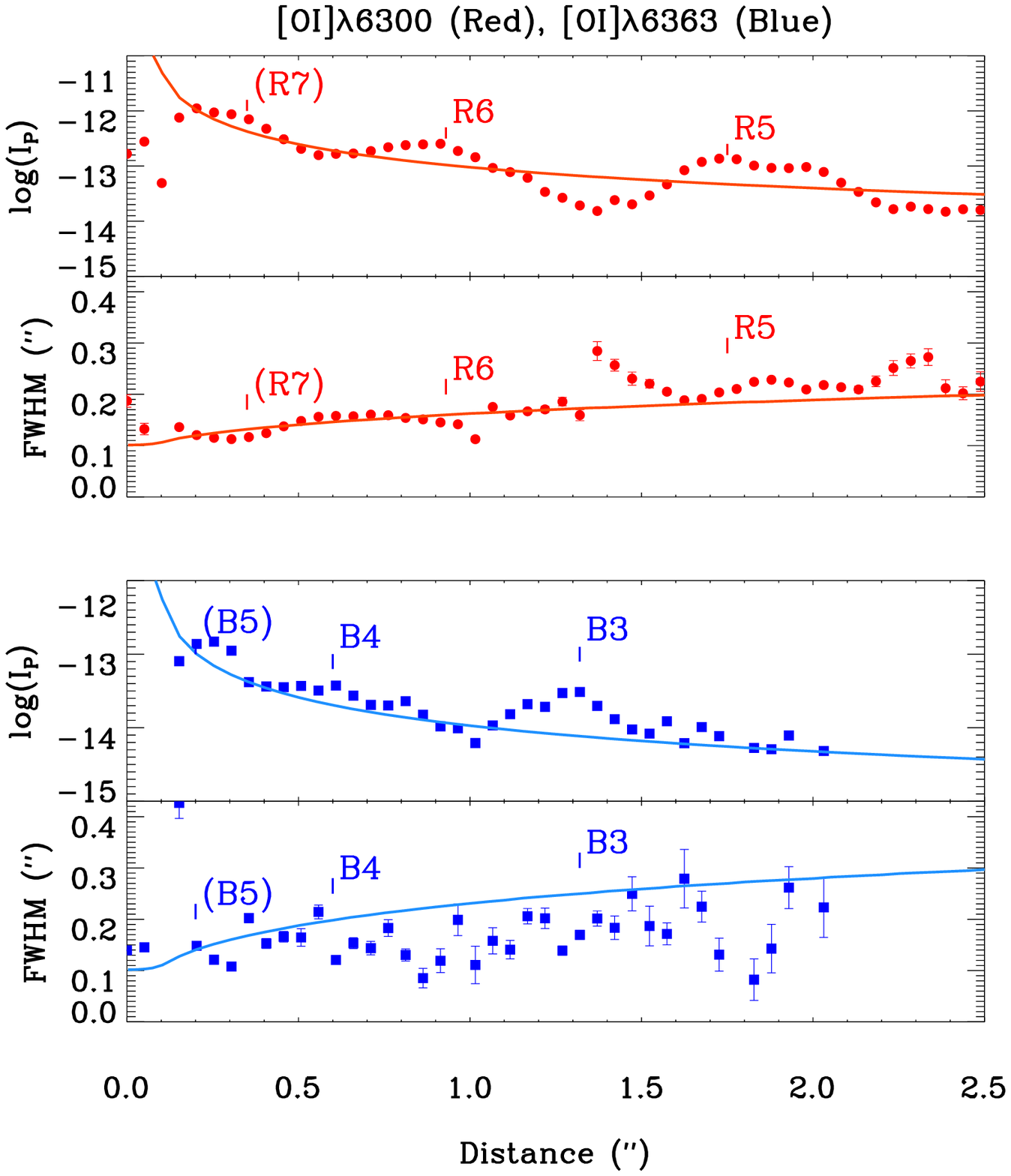}{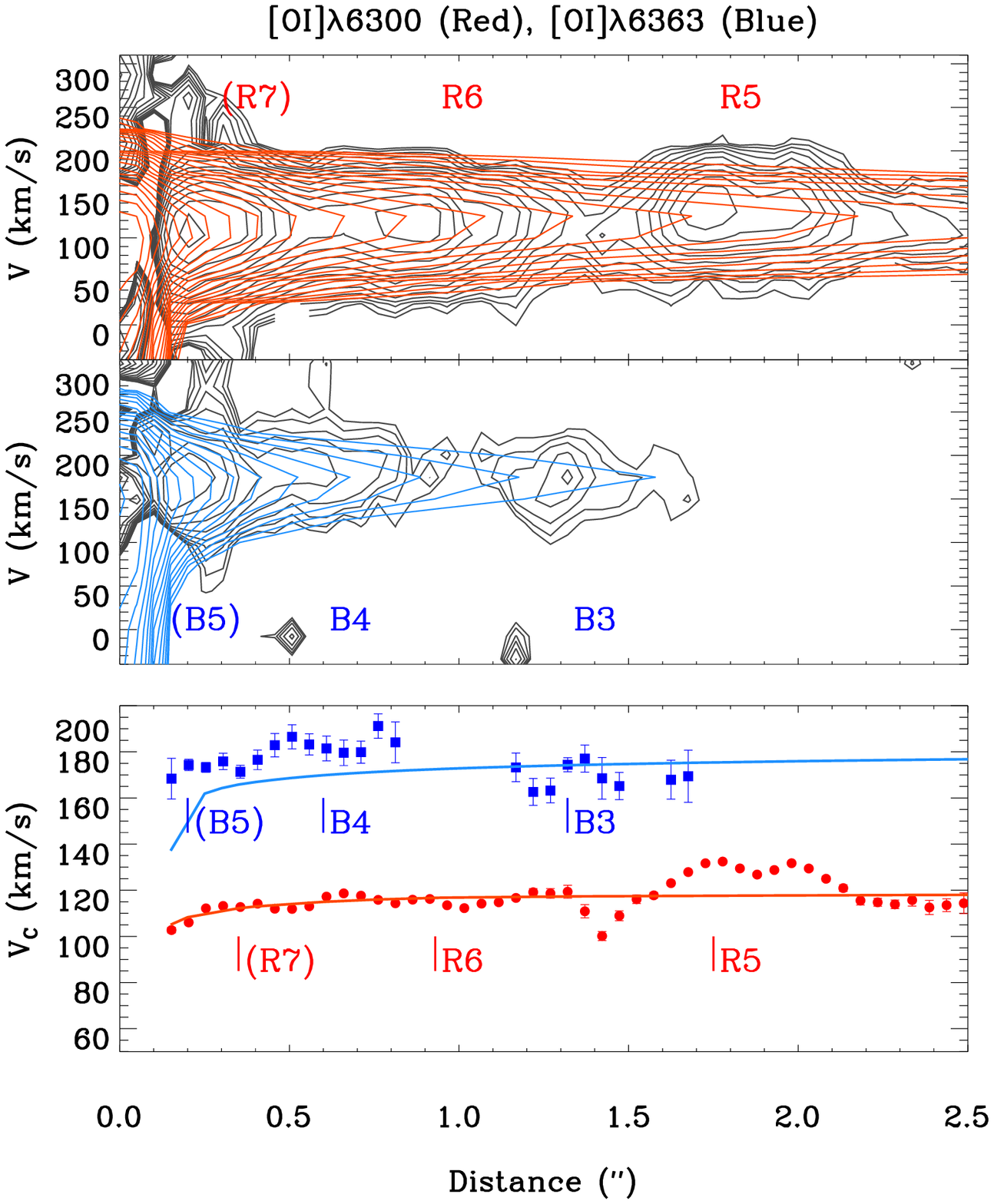}
\caption{Images and spectra as observed and theoretical data are compared on top of each other in [O {\sc i}] $\lambda6300$ (for the red jet) and [O {\sc i}] $\lambda6363$ (for the blue jet). {\it Left:} Variations of the peak intensity and the full-width-half-maxium (FWHM) are plotted along the axes of jets, with their gaussian fits from the synthetic images (thick lines). The red jet is shown in red and the blue jet is shown in blue. Observed data points are shown in red filled circles for the red jet and in blue filled squares for the blue jet.  {\it Right:} The position-velocity diagrams extracted from the observed (black contours) and synthetic (blue and red contours) spectra for the red and the blue jets in the upper panel. The Gaussian-fitted velocity centroids from the observed (blue and red filled symbols) and synthetic (blue and red thick lines) spectra are plotted in the lower panel, showing the distinct velocity ratios of about 1.6.}
\label{OI}
\end{figure}

\begin{figure}
\plottwo{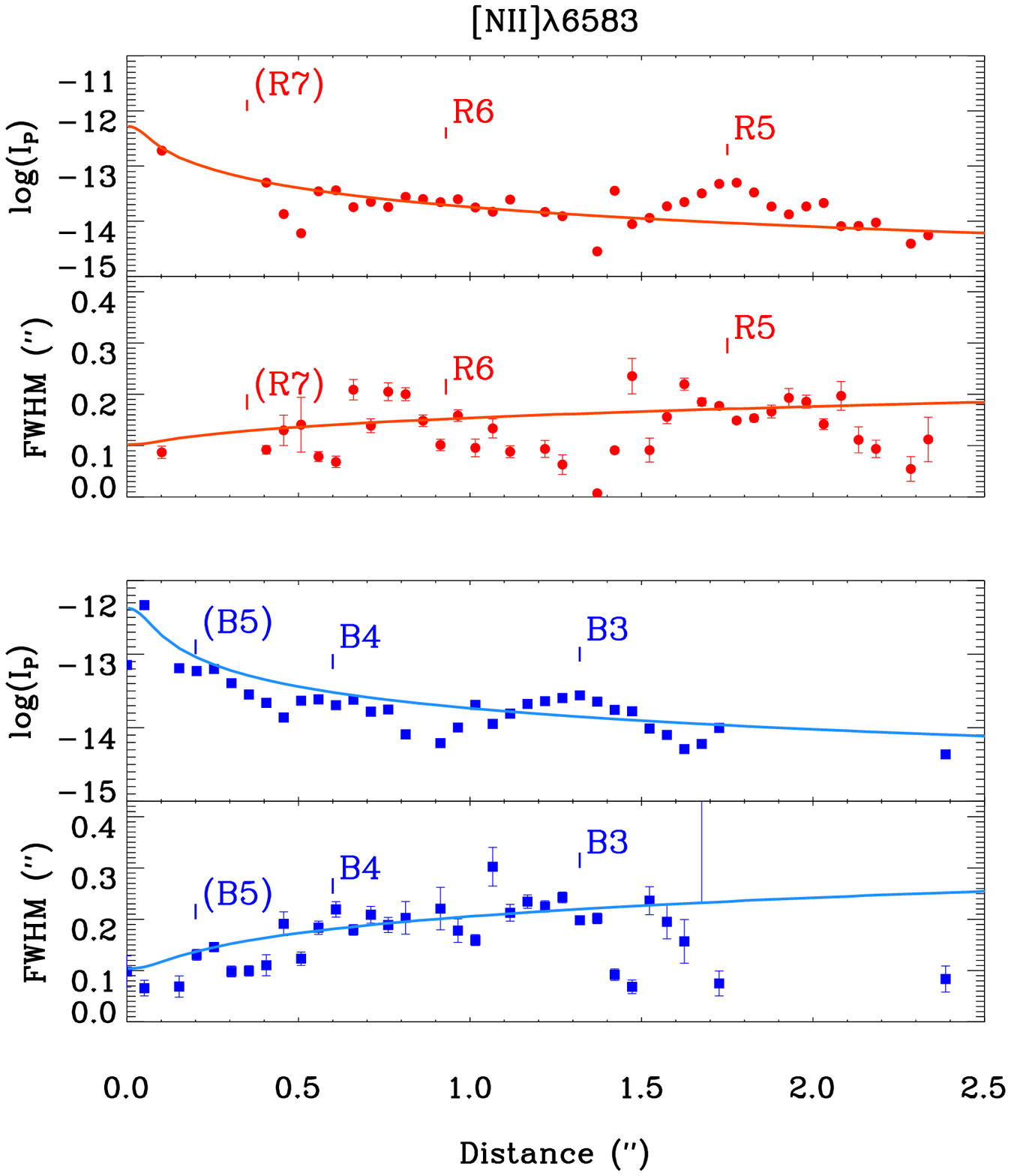}{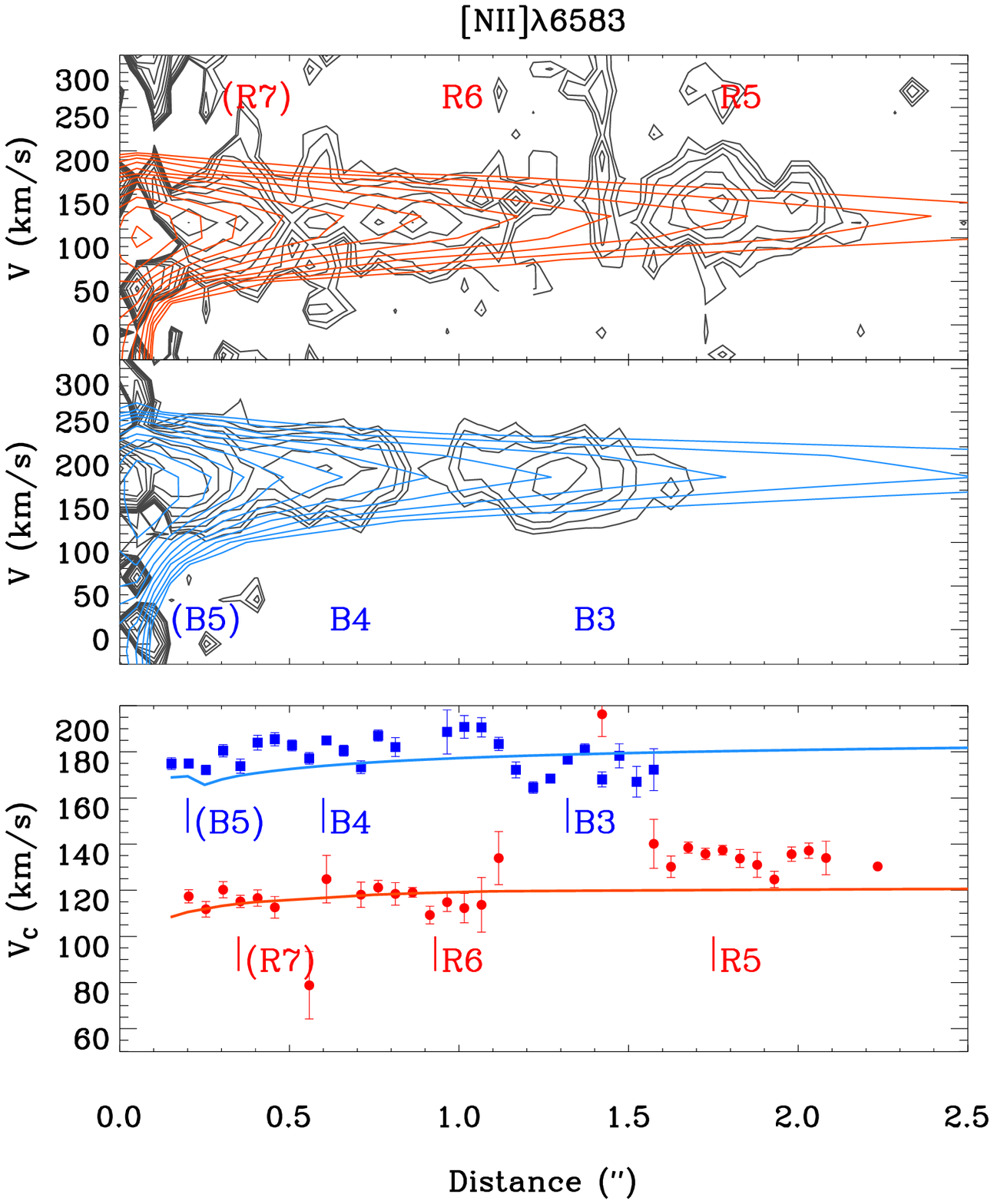}
\caption{Images and spectra as observed and theoretical data are compared on top of each other in [N {\sc ii}] $\lambda6583$. The meaning of the symbols is the same as that in Figure \ref{OI}.}
\label{NII}
\end{figure}

\begin{figure}
\plottwo{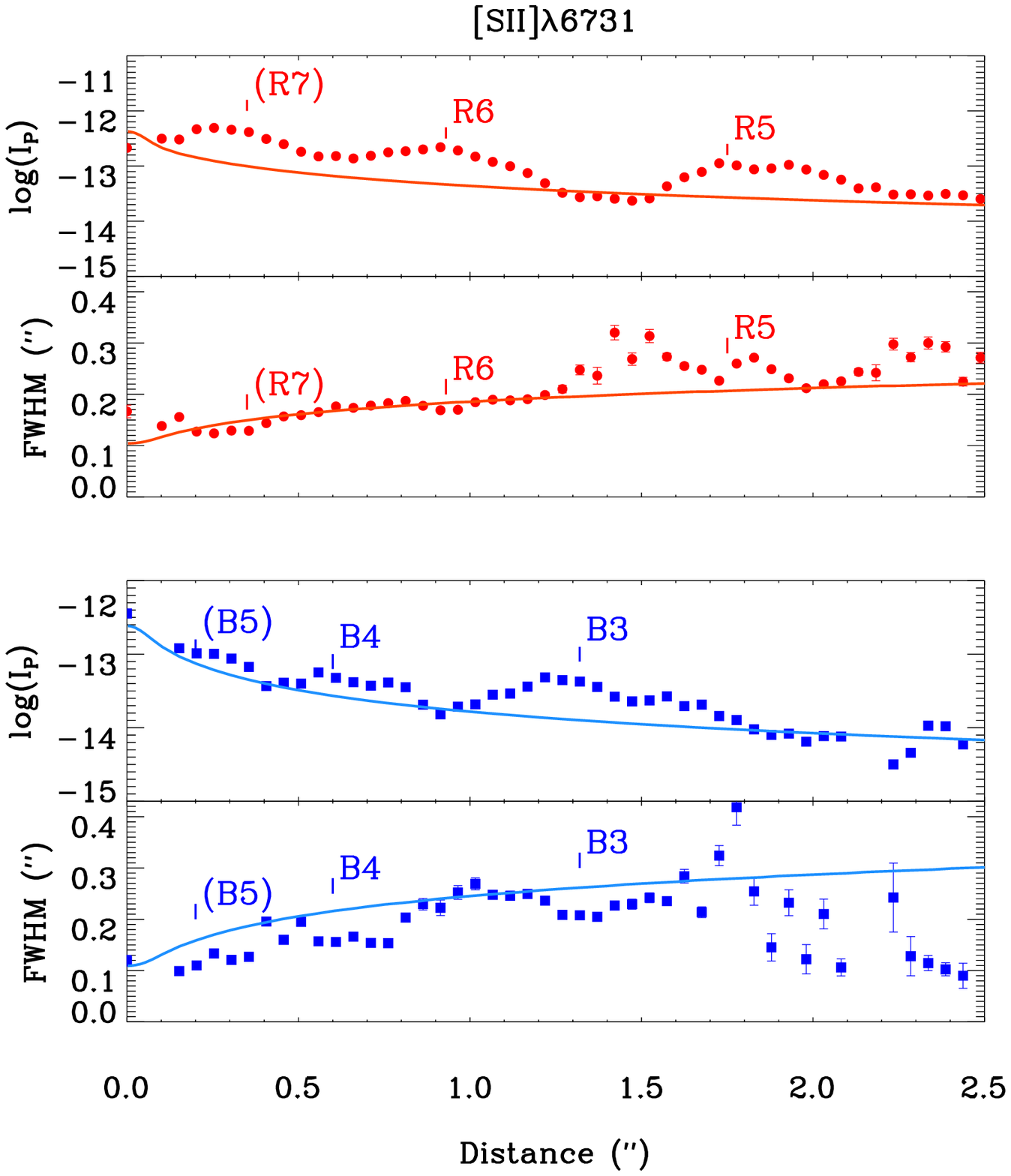}{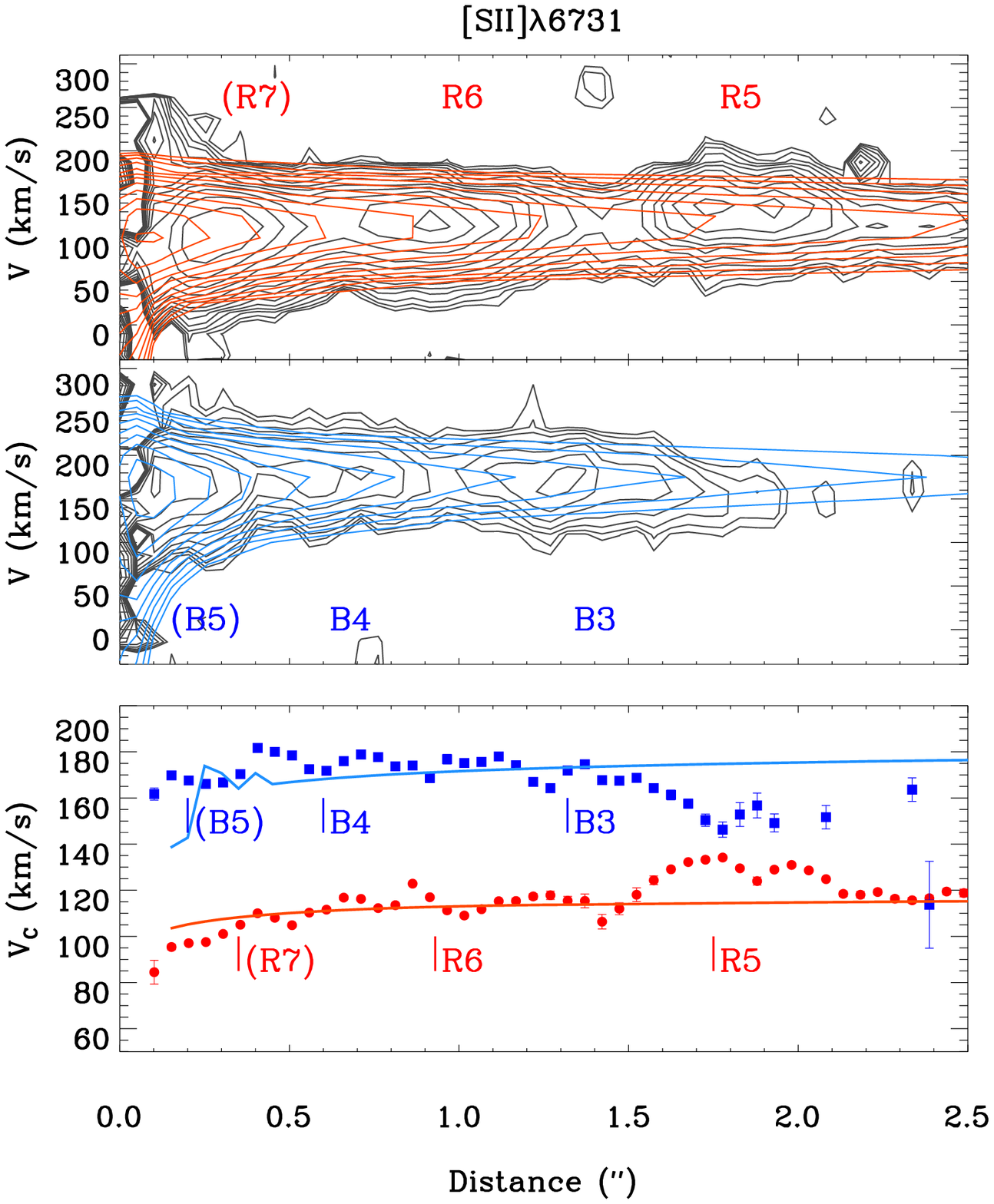}
\caption{Images and spectra as observed and theoretical data are compared on top of each other in [S {\sc ii}] $\lambda6731$. The meaning of the symbols is the same as that in Figure \ref{OI}.} 
\label{SII}
\end{figure}

\begin{figure}
\plotone{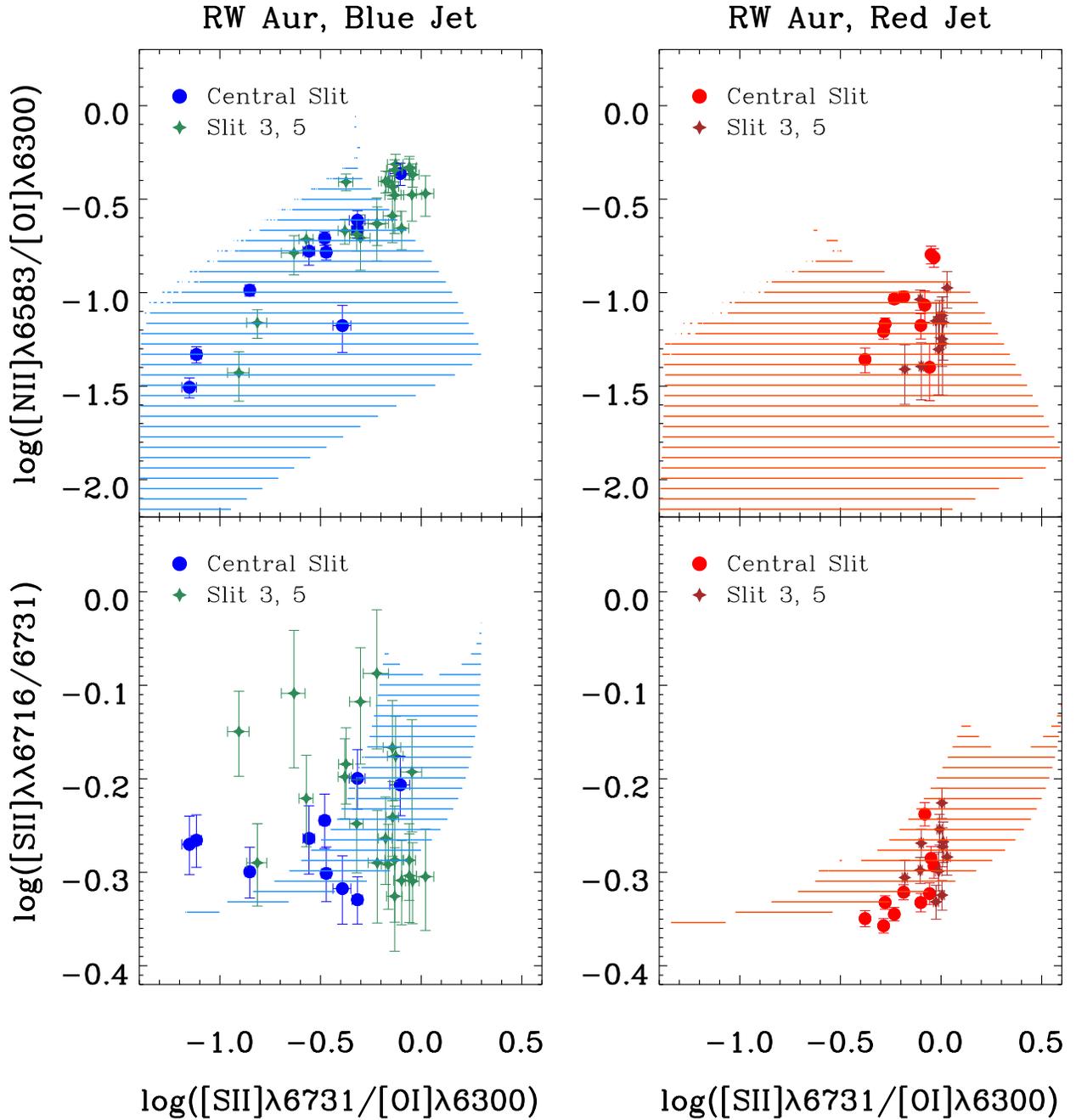}
\caption{Line ratios calculated from the blue jet and the red jet models (shaded areas) overlaid on the observed data (discrete symbols). The discrete symbols and shaded areas have the same spatial coverage in $x$ (0\farcs12, roughly corresponding to the central three slits) and $y$ (0\farcs95 for the blue jet and 0\farcs65 for the red jet).}
\label{CompareLR}
\end{figure}

We follow SGSL in finding the optimal parameters for mass-loss rates $\dot{M}_{\rm w}$, X-ray luminosities $L_{\rm X}$ and mechanical heating coefficient $\alpha_h$.  During the parameter searches, the ratio of X-ray luminosity to mass-loss rate, $L_{\rm X}/\dot{M}_{\rm w}$ is set to be $3\times10^{13}$ erg g$^{-1}$, as in SGSL, SLGS, and SGLL. From parameters adopted in the fiducial case of SGSL for both winds in the blue and red hemispheres, $\dot{M}_{\rm w}$, $L_{\rm X}$, and $\alpha_h$ are individually varied on each side to find the optimal matches for the intensity profiles and line ratios. The optimal sets of parameters are 
given in Table \ref{RunPars}, which capture the overall features reasonably well. Although the emission is knotty and the profiles and velocities have variations and scatter, clear average trends can be clearly identified for individual lines along 
the jet axis. For individual mass-loss rates of $3\times 10^{-8} M_\odot$\,yr$^{-1}$ in blue and $5\times 10^{-8}M_\odot$\,yr$^{-1}$ in red, and $\alpha_h=0.0010$ for the blue and $\alpha_h=0.0015$ for the red, the physical properties and emission 
profiles are illustrated in Figures \ref{NeVpPlot} to \ref{CompareLR}.

The physical scale in Figures \ref{NeVpPlot} and \ref{XeTPlot} covers the transverse 
and longitudinal dimensions of the seven slits: 50 AU in horizontal distance $\varpi$ 
and 150 AU in length $z$ along the jet axis. Figure \ref{NeVpPlot} shows the velocity 
contours on top of the electron density profiles, and Figure \ref{XeTPlot} shows the 
distributions of electron fraction and temperature. In Figure \ref{NeVpPlot}, 
the electron density is also cylindrically stratified as the underlying density as 
described in SSG from within 50 AU out to larger distances. Near the base of the jet, it 
casts an opening of a conic shape in the images as pointed out in SGSL. The velocity increases 
toward the axes on both sides of the jet, where the flows are denser and faster, until reaching 
the terminal velocities (\citealt{XWindV}; SSG). In Figure \ref{XeTPlot}, electron fraction 
and temperature follow similar behaviors to those in SGSL. Both $x_e$ and $T$ increase 
toward the inner boundaries of the wind, and peak within 5 AU from the launch point. 
The blue jet has a higher level of electron fraction and temperature than the red jet, 
supporting the trends observed from the line ratio diagnosis diagrams.

We show comparisons in PV diagrams, intensity profiles, image full-widths at half maximum 
(FWHM), and velocity centroids in [O {\sc i}]\footnote{We remind the reader that different [O {\sc i}] doublet lines are used for the red and blue jets: $\lambda6300$ for the red and $\lambda6363$ for the blue, due to the cut off of the blue wing of $\lambda$6300 at instrument level.}, [N {\sc ii}] $\lambda6583$, and [S {\sc ii}] $\lambda6731$ in Figures \ref{OI} to \ref{SII}. 
On the left, the profiles of intensity peaks and their FWHMs along the jet are shown for 
the inner 2\farcs5 from the source. On the right, the contours of the synthetic emission spectra 
(red for the red jet, and blue for the blue jet), convolved with the instrumentation profile, 
are overlaid on top of the observed PV diagrams. The lowest contours and 
contour intervals of both spectra are identical to those of Figure \ref{ObsPV}. Velocity 
centroids from both the observed and synthetic spectra are shown at the STIS velocity 
resolution of $\sim 65$ \kmps. The averaged velocities
at centroids of the synthetic spectra are $\sim -170$ \kmps\ for the blue and $\sim +110$ 
\kmps\ for the red, which reconfirm the velocity ratio of $\sim$ 1.6, and the trend of 
slightly increasing velocities along the axes. 
In the case of [N {\sc ii}] from the blue jet, emission at the 
stellar position is missing. However, we can not assess whether this is due to the nature 
of weaker emission from the blue jet or other reasons.

The line intensities and image widths of the synthetic images cover the overall trends in the observed ones. The trends delineated by the theoretical curves capture the major features well up to locations where the observed knots become too faint. For the redshifted jet, the model profiles go through the average of the knot peaks and off-knot regions; in the blueshifted jet, the same holds for the intensity profiles, 
although for the image widths and velocity centroids, the curves go through more scattered points except for the [S {\sc ii}] $\lambda 6731$. The image widths increase gradually with distance as the overall intensity profiles decrease gradually along the axes. This is expected as the electron fraction overall follows a pattern of recombination pointed out by \citet{BHN96} and SGSL.
 
The similarities and differences in the cross line-ratio diagrams between the synthetic maps and the actual data reveal the degree to which the physical conditions are reproduced by the model parameters. Figure \ref{CompareLR} shows the line ratios between [N {\sc ii}] $\lambda6583$/[O {\sc i}] $\lambda6300$ and [S {\sc ii}] $\lambda6731$/[O {\sc i}] $\lambda6300$, and between [S {\sc ii}] $\lambda\lambda6716/6731$ and [S {\sc ii}] $\lambda6731$/[O {\sc i}] $\lambda6300$ overlaid by model coverage (in color shades). The shaded areas include the innermost 0\farcs95 of the blue jet and 0\farcs65 of the red jet, at a distance ratio approximately proportional to their respective velocities. In the transverse direction, the model points extend to about $\varpi=0\farcs12$, corresponding roughly to the spatial coverage of the central three slits. The shaded areas generated by the synthetic images encompass most of the observed data points (in discrete symbols), following similar trends, up to the tips of the knots, B4 and R7. The overall ratio of [N {\sc ii}] $\lambda6583$/[O {\sc i}] $\lambda6300$ is about twice as large in the blue jet as in the red jet. As shown in Figure \ref{XeTPlot}, the overall electron fraction of the blue wind is approximately twice as large as in the red wind. The spatial dependence of [S {\sc ii}] $\lambda6731$/[O {\sc i}] $\lambda6300$ varies with the differences in critical densities in [S {\sc ii}] $\lambda6731$ and in [O {\sc i}] $\lambda6300$, and the [S {\sc ii}] doublet ratio varies across the jet as $n_e$ decreases faster laterally than longitudinally in the model. The combination of $(x_e,T)$ in Figure \ref{XeTPlot} following our selection criteria of the solutions indeed captures the overall sense of differences and asymmetry found on the opposite sides of the jet. The differences in excitation conditions within the opposite jet components differentiate the distribution of data points on the respective line ratio diagrams. 

Our synthetic points are distributed over a wider area than the observed points. This is because a large area is covered by fainter pixels on the computational domain.
This is especially noticeable for the brighter red jet. The model points come from the equivalent area covered by the central three slits as small pixels.  Emission from these theoretical pixels is excited ideally by the model approach outlined in \S\,2.2. As each of the theoretical pixels contributes to the model line-ratio points, there may not be corresponding points from the observed data. This is because the actual jets are knottier than the model ones and some of the jet volume in real space is not sufficiently excited as in the model counterpart. This explains why the model naturally predicts a broader range of values in line ratios while the observed range is narrower because observations are naturally weighted toward the regions with strong emission. On the other hand, the observed data for the blue jet have much higher uncertainties because the jet itself is weaker in emission but faster in flow speed, which may result in a wider spread in physical conditions to begin with.  However, a rough trend can still be traced as in the left panel of Figure \ref{ObsLR}, when all the slits are added to increase the signal-to-noise ratio. In this case, an overall match in the global trends are what we are seeking in the process.  We also note here that model points from the equivalent of three central slits have already covered most of the regions where emission lines are excited. Adding the emitting area of two more slits outside of the central area does not produce significant increase of the emission. Rather, the additional emitting area will mostly contribute to points with lower [O {\sc i}] $\lambda6300$ emission, resulting in higher [S {\sc ii}] $\lambda6731$/[O {\sc i}] $\lambda6300$ and [N {\sc ii}] $\lambda6583$/[O {\sc i}] $\lambda6300$ ratios, and [S {\sc ii}] doublet ratios approaching unity. These points may not be sensitively sampled compared to the brighter points represented by the observed data.

\section{Discussion}

\subsection{General Features}

We have attempted to re-analyze the set of archived {\it HST}/STIS data previously analyzed and published in \citet{Woi02} and \citet{Mel09}. Similar results were reached in general, although with some noticeable differences in terms of inferred physical quantities. Our methods of data analysis differ from those adopted in \citet{Mel09}, but are mostly similar to those adopted in \citet{Woi02}. \citet{Mel09} included further noise subtraction and correction of $A_V\approx0.4$ throughout the jet. The remaining noise in our data could introduce more scatter for the line ratios in the blue jet, but the overall values are consistent since the line ratios in adjacent wavelengths can mitigate the effect of extinction to $\sim 10\%$ \citep{BE99}. They also adopted a higher radial velocity for RW Aur A ($+23.5$ \kmps) than the averaged value ($+16.0$ \kmps). As expected, the line properties are nearly identical apart from the systematic velocity shift. On the results from archived data, the overall properties one would infer in general from the opposite sides of the jet are not altered by different approaches and assumptions.

The physical conditions derived from the different diagnostic analyses, rather than the methods of data analysis, distinguish the various works based on the same set of data. \citet{Mel09} derived the physical quantities using techniques
developed by \citet{BE99} for the individual jet without making {\it a priori}
assumptions of the fainter blue jet as in \citet{Woi02}. \citet{Mel09} reached the
conclusion that the blue jet is less dense and more excited than the red jet. The abundances of the atomic and ionic species involved are somewhat different, which may contribute to
theoretical line intensities and ratios through the combination of electron density, ionization fraction,
and temperature. Comparing the physical conditions derived from our self-consistent approaches (Figures
\ref{NeVpPlot} and \ref{XeTPlot}) and those derived from diagnostic mapping by employing the BE
technique (Table 1 of \citealt{Mel09}), we note differences in values and relative trends.
For example, electron densities derived from the BE technique are lower in the red jet by 20\%. The faintness of the blue jet introduces larger uncertainties in line ratios and in the derived electron densities. Values derived from the [S {\sc ii}] doublet may also represent a lower limit due to its low critical density ($\sim 2.5\times10^3$ cm$^{-3}$). It can be seen in Figure \ref{NeVpPlot} that values for both jets are similar in the densest region, but they decrease faster in the cylindrical
distance $\varpi$ in the blue. From the ionization fraction and temperature, both works concluded that the blue jet
is the more heated and ionized, and the temperatures derived from the BE technique are typically 50\%
higher than those found in this work. The higher temperature may compensate for the lower inferred electron densities for the observed line intensities. The re-heating of the jet close to knot positions may raise the values locally. The derived ionization fraction from \citet{Mel09} is in line with our values at the innermost streamlines where X-ray irradiation is the strongest (Figure \ref{XeTPlot}). The hydrogen density that is derived by dividing the electron density with the electron fraction, may weight toward the lower value, leading to mass-loss rates that could be lower by an order of magnitude (see below).

Physical properties such as the mass-loss rates inferred for the opposite sides of the jet are the best tests for the dynamical models and diagnostic analyses developed and integrated for the models.  
Our best-fit mass-loss rate for the red jet, is in line with that found in \citet{BHN96} using line ratios, and those found in \citet{HEG95} and \citet{WH04} using forbidden line intensities. These values are of the same order as those in \citet{Woi02}, but one order of magnitude larger than those derived by \citet{Mel09}.  
The small value in \citet{Mel09} lies in their consideration of ``intrinsic'' image widths. Although the jet radii appear more accurate, the mass loss carried by the associated wide-angle wind is not taken into account. It is also possible that the electron densities derived from [S {\sc ii}] doublet may represent a lower limit.
On the other hand, the imbalance of momentum deduced from the ratio of mass-loss rates between the two sides of the jet is larger in \citet{Woi02}.
Specifically, \citet{Woi02} obtained line-of-sight velocity, $v_j$, and jet width, $r_j$, from the same dataset, and derived electron densities $n_e$ from the [S {\sc ii}] line ratio. They inferred a mass-loss rate $\dot{M}_{\rm w} = \mu m_{\rm H}(n_e/x_e)\pi v_j r^2_j / \cos i$, where electron fraction $x_e$ of the red jet was adopted from \citet{DCL02}, and the values in the blue jet were assumed to be similar to those in the red jet. They also assumed similar [N {\sc ii}] $\lambda6583$/[O {\sc i}] $\lambda6300$ ratios for both sides, which led to an estimate of a mass-loss rate ratio $\sim 2$ (blue to red). Similarly, a ratio of $\sim 3$ is derived for the momentum. \citet{Woi02} interpreted this as a temporary phenomenon during accretion since there was no recoil velocity detected.  
Yet the discrepancy may be resolved if the adopted $x_e$ is 2 to 3 times higher in the blue jet, then the ratio of momentum rates from the two sides would be, $\propto v^2_j/x_e\approx 1$, after correcting the ratio of electron fraction. 
In addition, their analysis of the emitted fluxes gave mass-loss rates that decrease by more than one order of magnitude within a distance of 2\arcsec, which they interpreted as a decrease in mass loss due to sideway push at internal working surfaces. Although they also acknowledged the possibility of a decrease in electron fraction and a reduction of excitation, they left the unbalanced mass-loss and momentum rates unsolved.  
Our method self-consistently calculates the physical properties that best match the overall line intensities and line ratios, not biased toward only the inner dense part, leading to a better estimate of the mass-loss rates and a general balance of linear momentum.

The ``excess'' redshifted emission at the stellar position on the approaching side is difficult to assess due to subtraction artifacts of the stellar continuum. \citet{Pyo06} and \citet{HarHill09} pointed out the high-velocity features of the RW Aur jet in [Fe {\sc ii}] $\lambda$1.644 \micron\ at a velocity resolution $\lesssim 30$ \kmps: narrow line widths at the jet proper, symmetric profiles at each position, and increasing line widths toward the source, in general supports the predicted PV diagrams shown in SSG, rather than the disk-wind models. On the other hand, they also commented that the original PV diagrams for an approaching X-wind jet shown in SSG perhaps carry some ``excess'' redshifted emission extending over the stellar position toward the counter side. \citet{HarHill09} suggested stronger heating (or re-heating) at larger distance may resolve the discrepancy. The wide-angle wind would inevitably possess streamlines that contribute to the redshifted line-of-sight velocity, if the system is not strictly pole-on. We followed SSG and SGSL for the calculations of the line emissivities. The so-called ``excess'' is less pronounced in our maps generated for RW Aur A parameters with self-consistent excitation conditions implemented, when compared to the original PV diagrams in SSG that were produced with uniform temperature and ionization throughout the wind. This might suggest that the inner streamlines in our system might have been slightly more excited than the wind case inside the RW Aur A system. With 0\farcs1 spatial resolution of {\it HST}, the emission is restricted to the stellar position within one resolution element, and this area is strongly affected by the artifact introduced by stellar baseline subtraction in a rectified two-dimensional STIS spectra \citep[see also][]{Mel09}. It is very difficult to assess the actual ``excess'' of the redshifted emission.

We have demonstrated the existence of a pair of X-wind solutions that could capture the major features of asymmetry in the RW Aur A system. The asymmetric properties reported in the literature and the archival {\it HST}/STIS data are confirmed by our analyses. Mass-loss rates that are required to best match the overall trends in the emission profiles and line ratios are also asymmetric: 
$\dot{M}_B=3\times 10^{-8}\,M_\odot$\,yr$^{-1}$ for the blue, and 
$\dot{M}_R=5\times 10^{-8}\,M_\odot$\,yr$^{-1}$ for the red. These two mass-loss rates 
give a ratio of $\dot{M}_B/\dot{M}_R\approx 0.6$. Multiplied by the asymmetric terminal velocity 
ratio, $\bar{v}_B/\bar{v}_R\approx 1.6$, we reach a ratio of linear momenta $\dot{M}_B\,\bar{v}_B/\dot{M}_R\,\bar{v}_R\approx 1$. This suggests that momentum ejected into the opposite hemispheres is indeed balanced. This is in contrast to the apparent dilemma of unbalanced 
momenta carried by the two jets from the velocity asymmetry reported in \citet{Woi02} and \citet{Mel09}. Despite the asymmetry in velocity, line intensities, and mass-loss rates, the system is on average balanced for the most important physical 
quantity.

The apparent unequal mass-loading and the uneven terminal velocities in the northern 
and southern hemispheres of RW Aur A do not break the law of conservation for linear 
momenta and magnetic fluxes contributed in the wind in our study. These major physical 
quantities are indeed conserved in the average sense across the opposite jets. 
From the analysis of emission lines originating from the jet alone, the conservation is an integral part of the original scenario and framework laid out in the original works of the X-wind 
models \citep{XWindI,XWindII,XWindV}, and there is no need to invoke asymmetry beyond 
the immediate wind proper. 
There is also no obvious detectable precession or signatures 
from precession of the jet that would suggest a non-alignment of the magnetic axis 
with the rotation axis on the physical and time scales of the existing datasets. 
However, we can not rule out other possibilities from the star 
itself or processes of the star-disk interaction. On the other hand, if the timescales 
of variation are short, effects of asymmetry originating deep in the star system might 
have been averaged over the rotation periods and long flow time so the visual effects 
on the large scale are small. Further clues along these lines supported by other 
observational evidence will be discussed in a follow-up investigation.

\subsection{Further Constraints}

In this work, we demonstrated the possibilities of finding a match between the 
X-wind models and the {\it HST}/STIS spectra of the RW Aur A jet, and discussed additional 
properties that were inferred from such a matching process. Although the existence 
of such a set of solutions does not constrain a unique launch point $R_{\rm x}$, the overall 
procedure suggests that general features of the X-wind models indeed match those obtained from the data analysis. In particular, the mass-loss rates and excitation properties not only fall nicely within the general range 
suggested by the data but also solve the puzzling asymmetry in the physical properties of the system. Despite such ambiguity, the general results demonstrated in this work should exist in any model that is to be selected should a 
different combination using another $R_{\rm x}$ provides a good fit with the detailed properties. 
The differences should be within a factor of two of what we learned in this work.

Current available observational data on the RW Aur A jet alone do not seem to be 
able to resolve the ambiguity. Observational clues or constraints derived near the 
base of the jet related to the inner truncation radii of the disk, or some other 
information inferred from the permitted lines originating in the accretion funnels 
may help resolve the uncertainties. Inner truncation radii of the RW Aur A disk 
from near-infrared interferometry \citep{Ake05,Eis07} have a 
range of 0.08 to 0.2 AU, consistent with fits of models with dust sublimation temperature 
between 1000 to 1500 K. The gas disk may extend to 0.02 to 0.05 AU 
\citep{Eis09} inside of the dust disk although the actual values may be dependent on the underlying disk models adopted. These are all well within 
the considerations of choices for $R_{\rm x}$ given in this paper (Tables \ref{SetI} and 
\ref{SetII}). 

Recent Zeeman-splitting measurement of optical emission lines can help to qualitatively infer the magnetic structure of the interaction region between the star and the disk \citep{DLC12_Bz}. The averaged longitudinal magnetic field strength $B_z$ inferred from the narrow component of He~{\sc i}~$\lambda5876$ was found to change polarity within one rotation period assuming $P=5.576$ days. It is consistent with a model proposed in \citet{Pet01} that the variation of the lines is due to two hot spots on opposite poles of the star. The co-rotation radius corresponding to this period is $\sim 0.07$ AU assuming $M_\ast=M_\odot$. While this is still within our considerations, it may also be possible that the disk truncation radius ($\sim R_{\rm x}$) does not always coincide with the co-rotation radius. In the picture of a fluctuating X-wind \citep{Shu97Sci}, $\Omega_\ast \neq \Omega_{\rm x}$, the magnetic energy released would be proportional to the difference in angular frequency, and part of it can be in X-rays. In this case, we know that the X-ray activities may affect the initial ionization of the flow and properties of the lines. Simultaneous monitoring in both the X-rays and photospheric and forbidden emission will be desirable to understand the star-disk interaction in the RW Aur A system.

\section{Summary}

We have presented detailed analyses and comparisons between the {\it HST}/STIS archival data and synthetic spectra from the X-wind models for the bipolar jet of RW Aur A. We analyzed properties of optical forbidden lines [O {\sc i}], [N {\sc ii}], and [S {\sc ii}] through PV diagrams, images, and line ratios. PV diagrams of all the lines show similar kinematic features, and an average blue-to-red velocity ratio of $\sim 1.6$. Spectra of high spatial resolution enable diagnostics for both jets through the line ratios.
We were able to identify general properties related to the asymmetry of physical conditions derived through the line ratios, and applied them as constraints for model comparisons.

The diagnostic approach developed in SGSL was extended and applied to RW Aur A, using parameters for the stellar mass, radius, systemic velocity, and inclination angles of the jet and disk. Within observational uncertainties, we could select a set of stellar parameters to demonstrate a sample fitting process for illustration. Parameters were searched for mass-loss rates, X-ray luminosities, and mechanical heating starting from the fiducial models adopted in \citet{SGSL,SLGS,SGLL} for a pair of X-wind solutions that were selected by matching the velocity ratio. The combined models not only possess the feature of asymmetry in jet speeds, but also simultaneously meet the overall agreement in line intensities, image widths, and line ratios.

Our interpretation of the RW Aur A system offers a different perspective to the driving of the asymmetric jets. Since its discovery, the asymmetry in the opposite sides of the bipolar jet has intrigued astronomers for many years. Previous analysis of the same STIS dataset suggested a nearly equal mass-loss rate in the opposite jets themselves but with asymmetry in the ambient environment \citep{Mel09}. Through the demonstration of a sample fit with the X-wind models, we are able to interpret the asymmetry in velocity and intensity as a consequence of different mass loading on the opposite sides of the disk while the system adjusting itself to conserve linear momentum. Moreover, our analysis suggests the jet and wide-angle wind system as a whole is one entity, since 
observational interpretation through the optical lines alone would not recover the entirety of mass loss inside the hemisphere.

In this work we demonstrate that the framework of X-winds can accommodate many general features of asymmetric disk-jet systems such as RW Aur A. Analysis such as one presented here may be applied to other similar systems. Further constraints of the disk-jet system through high-dispersion spectral line diagnosis associated with the jet and accretion activities are strongly encouraged.

{\it Facility:} \facility{HST (STIS)}

\acknowledgments

We thank Francesca Bacciotti for numerous discussions at the very early phase of data reduction. We also thank Sylvie Cabrit and Catherine Dougados for their encouragement of taking on the tasks. We thank Tae-Soo Pyo for informative discussions on his Subaru observations. The authors are indebted to Gibor Basri, for pointing out the puzzling nature of the RW Aur system. {\it HST}/STIS spectra were obtained from the Multimission Archive at the Space Telescope Science Institute (MAST). STScI is operated by the Association of Universities for Research in Astronomy, Inc., under NASA contract NAS 5-26555. This work was supported by funds from the Theoretical Institute for Advanced Research in Astrophysics (TIARA), in part through the Academia Sinica and the Excellence Program of the National Science Council of Taiwan by NSC93-2752-M-001-001-PAE, NSC94-2752-M-001-001-PAE, NSC95-2752-M-001-008-PAE and NSC96-2752-M-001-001-PAE.


\clearpage

\begin{deluxetable}{lccccccccc}
\tablecolumns{10}
\tablecaption{Inferred Stellar Parameters}
\tablehead{\colhead{Spec Type} & \colhead{$\log T_{\rm eff}$} &
           \colhead{Band} & \colhead{$m_\lambda$} &
           \colhead{$r_\lambda$} & \colhead{$A_V$} &
           \colhead{$\log \frac{L_\ast}{L_\odot}$} &
           \colhead{$\frac{R_\ast}{R_\odot}$} & \colhead{$\frac{M_\ast}{M_\odot}$} & 
           \colhead{Reference}\\
           \colhead{(1)} & \colhead{(2)} & \colhead{(3)} & \colhead{(4)} & \colhead{(5)} & 
           \colhead{(6)} & \colhead{(7)} & \colhead{(8)} & \colhead{(9)} & \colhead{(10)}}
\startdata
K4    & 3.662       & $J$     &   8.54  & 0.18    & 2.2  & 0.36    & 2.4     & 0.85      & 1 \\
K4    & 3.662       & \nodata & \nodata & \nodata & 0.8  & \nodata & 2.0     & 1.0       & 2 \\
K3    & 3.672       & $J$     &   8.66  & 0.49    & 0.53 & 0.05    & 1.6     & 1.3       & 3 \\
K1/K3 & 3.706/3.675 & $I_C$   &   9.43  & $\sim0$ & 0.39 & 0.23    & 1.7/2.2 & 1.34/1.4  & 4
\enddata
\tablecomments{Description of columns \\
 \renewcommand{\labelenumi}{(\arabic{enumi}):}
 \begin{enumerate}
 \itemsep=-5pt
 \item Spectral type, with a typical uncertainty of one subclass; \\
 \item Effective temperature (in K) of the star for the spectral type in column (1) \citep{HSS94}; \\
 \item The band used to calculate the stellar luminosity; the central wavelengths for $J$ and $I_C$
       bands are 1.25 and 0.79 \micron, respectively; \\
 \item The magnitude (in mag) of the star measured with the band in column (3); \\
 \item Veiling (defined by the ratio between total continuum flux and stellar flux) at the 
       central wavelength of the band in column (3), derived from the veiling at 
       5700 \AA\ ($r_V$) by assuming $r_J = 0.1\,r_V$ and that $r_{I_C}$ is negligible; \\
 \item Visual extinction (in mag) used to calculate the extinction at the central wavelength of the band
       in column (3); \\
 \item Stellar luminosity (in solar luminosity) calculated from the magnitude in column (4), veiling in column (5),
       extinction in column (6), and the bolometric corrections from \citet{HSS94}; \\
 \item Stellar radius (in solar radius) derived from the effective temperature in column (2) and 
       the luminosity in column (7); \\
 \item Stellar mass (in solar mass) inferred from the theoretical evolutionary tracks on the H-R diagram.
 \end{enumerate}
}
\tablerefs{(1) \citet{HEG95}; (2) \citet{SBB00};
           (3) \citet{WLK01}; (4) \citet{WG01}.}
\label{starpars}
\end{deluxetable}

\begin{deluxetable}{lccccccc}
\tablecolumns{8}
\tablecaption{Radial Velocities in Literature and Resulting Velocity Ratios}
\tablehead{\colhead{$v_{\rm rad}$} & \colhead{min $v_{\rm rad}$} & \colhead{max $v_{\rm rad}$} &
           \colhead{Epochs} & \colhead{$\bar{v}_{\rm blue}$} &
           \colhead{$\bar{v}_{\rm red}$} & \colhead{$\bar{v}_{\rm blue}/\bar{v}_{\rm red}$} &
           \colhead{Reference} \\
           \colhead{(\kmps)} & \colhead{(\kmps)} & \colhead{(\kmps)} &
           \colhead{(year)} & \colhead{(\kmps)} & \colhead{(\kmps)} & \colhead{} & \colhead{}}
\startdata
$+14.0\pm4.9$  & $+7.3$ & $+22.4$ & $1981.9 - 1985.1$ & $-170.5$ & $+109.5$ & $1.56$ & 1 \\
$+11.0\pm10.0$ & $-2.0$ & $+22.0$ & $1996.9$          & $-167.5$ & $+112.5$ & $1.49$ & 2 \\
$+16.0\pm5.5$  & $+6.1$ & $+25.1$ & $1995.9 - 1999.9$ & $-172.5$ & $+107.5$ & $1.60$ & 3 \\
$+16.0\pm1.9$  & \nodata& \nodata & $2003.2$          & $-172.5$ & $+107.5$ & $1.60$ & 4 \\
$+23.5$        & \nodata& \nodata & $2000.9$          & $-180.0$ & $+100.0$ & $1.80$ & 5 \\
\enddata
\tablerefs{(1) \cite{HHSM86}; (2) \cite{SBB00}; (3) \cite{Pet01};
           (4) \cite{WH04}; (5) \cite{Woi05}.}
\label{RV}
\end{deluxetable}

\begin{deluxetable}{ccccc}
\tablewidth{0pc}
\tablecolumns{5}
\tablecaption{Sample Fitting Set I}
\tablehead{\colhead{$i$} & 
           \colhead{$v_{\rm w}({\rm blue})/v_{\rm x}$} & \colhead{$v_{\rm w}({\rm red})/v_{\rm x}$} &
           \colhead{$v_{\rm x}$} & \colhead{$R_{\rm x}$} \\
           \colhead{(\degr)} & \colhead{} & \colhead{} & \colhead{(\kmps)} & \colhead{(AU)}}
\startdata
45 & $3.208\rightarrow3.609$ & $2.005\rightarrow2.306$ & 70.534 & 0.1779 \\
50 & $3.208\rightarrow3.609$ & $2.005\rightarrow2.306$ & 77.591 & 0.1470 \\
55 & $3.208\rightarrow3.609$ & $2.005\rightarrow2.306$ & 86.953 & 0.1170 \\
60 & $3.208\rightarrow3.609$ & $2.005\rightarrow2.306$ & 99.751 & 0.0889 \\
\enddata
\label{SetI}
\end{deluxetable}

\begin{deluxetable}{ccccc}
\tablewidth{0pc}
\tablecolumns{5}
\tablecaption{Sample Fitting Set II}
\tablehead{\colhead{$i$} & 
           \colhead{$v_{\rm w}({\rm blue})/v_{\rm x}$} & \colhead{$v_{\rm w}({\rm red})/v_{\rm x}$} &
           \colhead{$v_{\rm x}$} & \colhead{$R_{\rm x}$} \\
           \colhead{(\degr)} & \colhead{} & \colhead{} & \colhead{(\kmps)} & \colhead{(AU)}}
\startdata
45 & $2.005\rightarrow2.306$ & $1.253\rightarrow1.473$ & 112.85 & 0.0695 \\
50 & $2.005\rightarrow2.306$ & $1.253\rightarrow1.473$ & 124.15 & 0.0574 \\
55 & $2.005\rightarrow2.306$ & $1.253\rightarrow1.473$ & 139.13 & 0.0457 \\
60 & $2.005\rightarrow2.306$ & $1.253\rightarrow1.473$ & 159.60 & 0.0347 \\
\enddata
\label{SetII}
\end{deluxetable}

\begin{deluxetable}{lcccc}
\tablewidth{0pc}
\tablecolumns{5}
\tablecaption{Adopted Parameters for the RW Aur General Set: ($M_\ast,R_\ast$) $=$ ($1M_\odot,2R_\odot$)}
\tablehead{\colhead{Properties} & \colhead{Symbol} & \colhead{Blue Jet} & \colhead{Red Jet} & \colhead{Unit}}
\startdata
Average terminal wind velocity& $\bar{v}_{\rm w}$ &  310  & 200  & \kmps \\
Inclination angle             & $i$               &  55   & 55  & \degr \\
Disk truncation radius        & $R_{\rm x}$       & 0.117 & 0.117  & AU \\
Mass-loss rate                & $\dot{M}_{\rm w}$ & $3\times10^{-8}$ & $5\times 10^{-8}$ & $M_\odot$\,yr$^{-1}$\\
X-ray luminosity              & $L_{\rm X}$       &  $6.0\times10^{31}$ & $9.9\times10^{31}$ & erg\,s$^{-1}$ \\
Mechanical heating coefficient& $\alpha_h$        &  $1.0\times10^{-3}$ & $1.5\times10^{-3}$ &  \\
\enddata
\label{RunPars}
\end{deluxetable}

\clearpage

\end{document}